\documentclass[preprint,nopreprintline]{elsarticle}

\usepackage{algorithm2e}
\usepackage{amsmath}
\usepackage{amssymb}
\usepackage{booktabs}
\usepackage{listings}
\usepackage{lineno}
\usepackage{url}

\newcommand{\tlmadjoint}{\texttt{tlm\_adjoint}}
\newcommand{\checkpointschedules}{\texttt{checkpoint\_schedules}}

\begin{document}


\begin{frontmatter}

\author[1]{James R. Maddison\corref{cor1}}
\ead{j.r.maddison@ed.ac.uk}

\cortext[cor1]{Corresponding author}
\affiliation[1]{organization={School of Mathematics and Maxwell Institute for Mathematical Sciences, The University of Edinburgh},
addressline={James Clerk Maxwell Building, Peter Guthrie Tait Road},
city={Edinburgh},
postcode={EH9 3FD},
country={United Kingdom}}

\title{Step-based checkpointing with high-level algorithmic differentiation}

\begin{abstract}
Automated code generation allows for a separation between the development of a model, expressed via a domain specific language, and lower level implementation details. Algorithmic differentiation can be applied symbolically at the level of the domain specific language, and the code generator reused to implement code required for an adjoint calculation. However the adjoint calculations are complicated by the well-known problem of storing or recomputing the forward data required by the adjoint, and different checkpointing strategies have been developed to tackle this problem. This article considers the combination of high-level algorithmic differentiation with step-based checkpointing schedules, with the primary application being for solvers of time-dependent partial differential equations. The focus is on algorithmic differentiation using a dynamically constructed record of forward operations, where the precise structure of the original forward calculation is unknown ahead-of-time. In addition, high-level approaches provide a simplified view of the model itself. This allows data required to restart and advance the forward, and data required to advance the adjoint, to be identified. The difference between the two types of data is here leveraged to implement checkpointing strategies with improved performance.
\end{abstract}


\end{frontmatter}

\noindent
Published article: `Step-based checkpointing with high-level algorithmic
differentiation' James R. Maddison, Journal of Computational Science 82,
102405, 2024, doi: 10.1016/j.jocs.2024.102405

\section{Introduction}

In \citet{farrell2013} a high-level approach for algorithmic differentiation is described, combining a view of a numerical code at the level of finite element discretized partial differential equations with automated code generation. The forward code is described using a domain specific language, the Unified Form Language \citep{alnaes2014}, and specific code necessary to assemble matrices and vectors is generated using the FEniCS code generator \citep{logg2012,alnaes2015}. The dolfin-adjoint library described in \citet{farrell2013}, and its successor pyadjoint \citep{mitusch2019}, use a variant of the standard operator overloading approach to algorithmic differentiation, intercepting forward calculations and building a record of problems solved as the forward calculation progresses. Symbolic differentiation is applied to build symbolic representations of components of the associated adjoint problem, and the code generator reused to construct implementations of code necessary for an adjoint calculation. With this approach the number of operations which need to be recorded is significantly reduced. For example the solution of a finite element discretized partial differential equation may appear as a single record.

Adjoint calculations proceed in a reverse causal sense to the original forward code, but also in general require access to forward solution data. This leads to the well-known problem of managing access to forward data for use by the adjoint -- forward data must either be stored for use by the adjoint, or it must be recomputed from other forward data in time for use by the adjoint. For large non-linear calculations it becomes infeasible to store all required forward data, as this will lead to available storage being exceeded. Checkpointing strategies have been developed to address this (see e.g. section 12.3 of \citet{griewank2008}), reducing storage demands at the cost of performing additional forward recalculation.

This article considers high-level algorithmic differentiation where the record of operations is constructed dynamically at runtime -- the difference between static and dynamic approaches is noted e.g. in \citet{baydin2018}. This article specifically considers the application of a dynamic approach to solvers of time-dependent partial differential equations, combined with the use of `step-based' checkpointing strategies. Solvers for these problems typically have a regular repeating structure, being logically divided into a sequence of `steps' which may correspond to timesteps in the solver. Crucially, however, while the steps in the forward calculation may be \emph{similar}, they cannot be assumed to be \emph{identical}. Breaks in structure may occur, for example, at the start or end of the calculation \citep{maddison2014}, but may in general occur at other arbitrary points. Checkpointing schedules must therefore be robust against arbitrary breaks in the structure. If the record of forward operations is constructed dynamically at runtime, then the checkpointing schedules must also be applicable when the detailed structure of the forward is unknown ahead-of-time. This article seeks to address these difficulties.

The revolve algorithm  \citep{griewank2000}, building on the approach of \citet{griewank1992}, provides optimal step-based checkpointing schedules for the case where the number of forward steps is known ahead of the calculation and where checkpoints store data required to restart the forward. Further approaches define schedules for the case where the number of steps is not known ahead of the forward calculation \citep[e.g.][]{wang2009,stumm2010}. Checkpointing schedules may also consider cases where there are different types of storage available \citep{stumm2009,aupy2016,schanen2016,herrmann2020}, or where compression is applied when storing checkpoints \citep{kukreja2019}. More general optimized checkpointing strategies, not making use of a regular step-based structure, may be challenging, noting for example that the problem of optimizing an adjoint calculation is itself NP-complete \citep{naumann2008}. However, more general approaches appear in the context of backpropagation in neural networks \citep[e.g.][]{jain2020,kirisame2021-preprint}.

It is important to note that only forward data actually used by the adjoint need be available -- either via storage or from  recomputation -- when performing an adjoint calculation. Specifically only the forward dependencies of the forward Jacobian matrix are required by the adjoint calculation -- see the definition of the \texttt{adjU} sets of variables identified in \citet{hascoet2005}. Such dependencies are here referred to as ``non-linear dependencies''. Further, for any given consecutive sequence of forward steps, non-linear dependency data for the steps can differ from the data required to restart and advance the forward over those steps. This difference, between what is here termed ``forward restart data'' and ``non-linear dependency data'', has recently been utilised in \citet{zhang2021} and \citet{zhang2023} in the context of multi-stage Runge-Kutta schemes to construct step-based checkpointing schedules with improved performance. The additional performance is achieved by permitting Runge-Kutta stage data to be stored in a checkpoint -- the stage data being non-linear dependencies when a Runge-Kutta method is applied to a non-linear ordinary differential equation.

When applying high-level algorithmic differentiation using the Unified Form Language, the symbolic representation of the forward allows non-linear dependencies to be identified. In this article this extra information is used to allow checkpointing schedules, which distinguish between forward restart data and non-linear dependency data, to be applied more generally, and in particular to be applied to models not making use of Runge-Kutta schemes, and to be applied even if the forward has arbitrary deviations from an otherwise regular repeating structure.

The revolve algorithm of \citet{griewank2000} is applied to high-level algorithmic differentiation with automated code generation in \citet{farrell2013} and \citet{maddison2019}. In this article a more general checkpointing schedule structure is described. The schedule explicitly incorporates the buffering of data in an ``intermediate storage'', so that that forward variables can be identified and computed by the forward as the calculation progresses, before later being stored in a checkpoint. This is an explicit form of checkpoint deferment -- previously used in \citet{maddison2019} -- and allows step-based checkpointing to be applied while also constructing the record of forward operations dynamically at runtime. The new schedule structure is sufficiently flexible to be applied to a number of existing approaches, including the revolve algorithm, the multistage approach of \citet{stumm2009}, the two-level mixed periodic/binomial approach described in \citet{pringle2016} and in the supporting information for \citet{goldberg2020}, and H-Revolve schedules \citep{herrmann2020}. The schedule further distinguishes between storage of forward restart data and storage of non-linear dependency data, allowing for the definition of schedules which make use of the difference between these two sets of dependencies, and which use of this difference for improved performance.

This article principally focuses on the application of checkpointing for adjoint calculations associated with models written using the Unified Form Language -- particularly FEniCS \citep{logg2012,alnaes2015} and Firedrake \citep{rathgeber2016}. The described approaches are implemented in Python in the \tlmadjoint{} library, and can be applied to any model which can be differentiated using \tlmadjoint{}, without further modification of code beyond the definition of forward steps and the schedule. The library \checkpointschedules{} has recently been developed using code from \tlmadjoint{}, and defines checkpointing schedules using an alternative set of schedule operations \citep{dolci2024}.

The article proceeds as follows. In section \ref{sect:forward_adjoint} details of forward and adjoint calculations, when viewed in terms of a high-level structure, are described. Section \ref{sect:schedule} describes a checkpointing schedule structure which can be applied in a high-level algorithmic differentiation approach, incorporating the use of an intermediate storage. This schedule structure is applied in section \ref{sect:sub_revolve} to implement a checkpointing schedule which makes use of the additional assumption that the sizes of forward restart data and single step non-linear dependency data are the same. The resulting schedule is a version of the CAMS-GEN algorithm of \citet{zhang2023} for $l = 1$ stage, but can be applied to a broader class of models. The article concludes in section \ref{sect:conclusions}.

\section{Forward and adjoint calculations}\label{sect:forward_adjoint}

A high-level algorithmic differentiation approach allows a calculation to be viewed in terms of a relatively small number of individually more complicated operations. Here the details of forward and adjoint calculations are described, so that different dependencies of the forward and adjoint can be identified.

\subsection{The computational graph}

Following the approach used by the pyadjoint library \citep{mitusch2018,mitusch2019}, we view the forward calculation in terms of a computational graph. See also \citet{abadi2016} for details regarding the use of ``dataflow graphs'' in TensorFlow.

The forward problem is divided into a number of operations, each of which computes values for one or more output variables using zero or more input parameters or variables. In a high-level approach these operations may correspond to the solution of a discrete partial differential equation, and may further be defined implicitly via the solution of a non-linear problem. The parameters or variables may, for example, consist of single scalars, vectors, or finite element discretized functions. The forward calculation may therefore be visualized via a computational graph which, after converting to static single-assignment form, is a directed acyclic graph.

Operations are further collected together into larger ``steps'', which may for example correspond to one or more timesteps in a time dependent numerical solver. Each operation (and similarly each step) is indexed, and the forward calculation performs the operations (and each step) in index order. As in \citet{griewank2000} it is assumed that an appropriate division of the forward into steps is provided, and forward or adjoint advances always occur over full steps.

As an example, the computational graph associated with a numerical solver for the barotropic vorticity equation on a beta plane is considered. The configuration corresponds to a time-dependent non-linear Stommel-Munk problem \citep[][chapter 14]{vallis2006}. The model is implemented using Firedrake \citep{rathgeber2016}, and \tlmadjoint{} \citep{maddison2019} is used to build the record of operations. The problem is discretized in space using $P_1$ continuous Lagrange finite elements, and in time with third order Adams-Bashforth, started with a forward Euler step followed by a second order Adams-Bashforth step. In visualizations of the computational graph for this example, $\psi$ corresponds to the stream function, $\zeta$ to the relative vorticity, $\psi_0$ to the initial stream function, $Q$ to the wind forcing term appearing in the vorticity equation, $\beta$ to the magnitude of the background planetary vorticity gradient, $r$ to the linear bottom drag parameter, and $\nu$ to the Laplacian viscosity coefficient. $F_0$, $F_1$, and $F_2$ correspond to the right-hand-side of the barotropic vorticity equation evaluated on different time steps, and $\zeta_\mathrm{prev}$ is an auxiliary variable used to store a previous value for the relative vorticity.

\begin{figure}\begin{centering}
\includegraphics[width=1.0\textwidth]{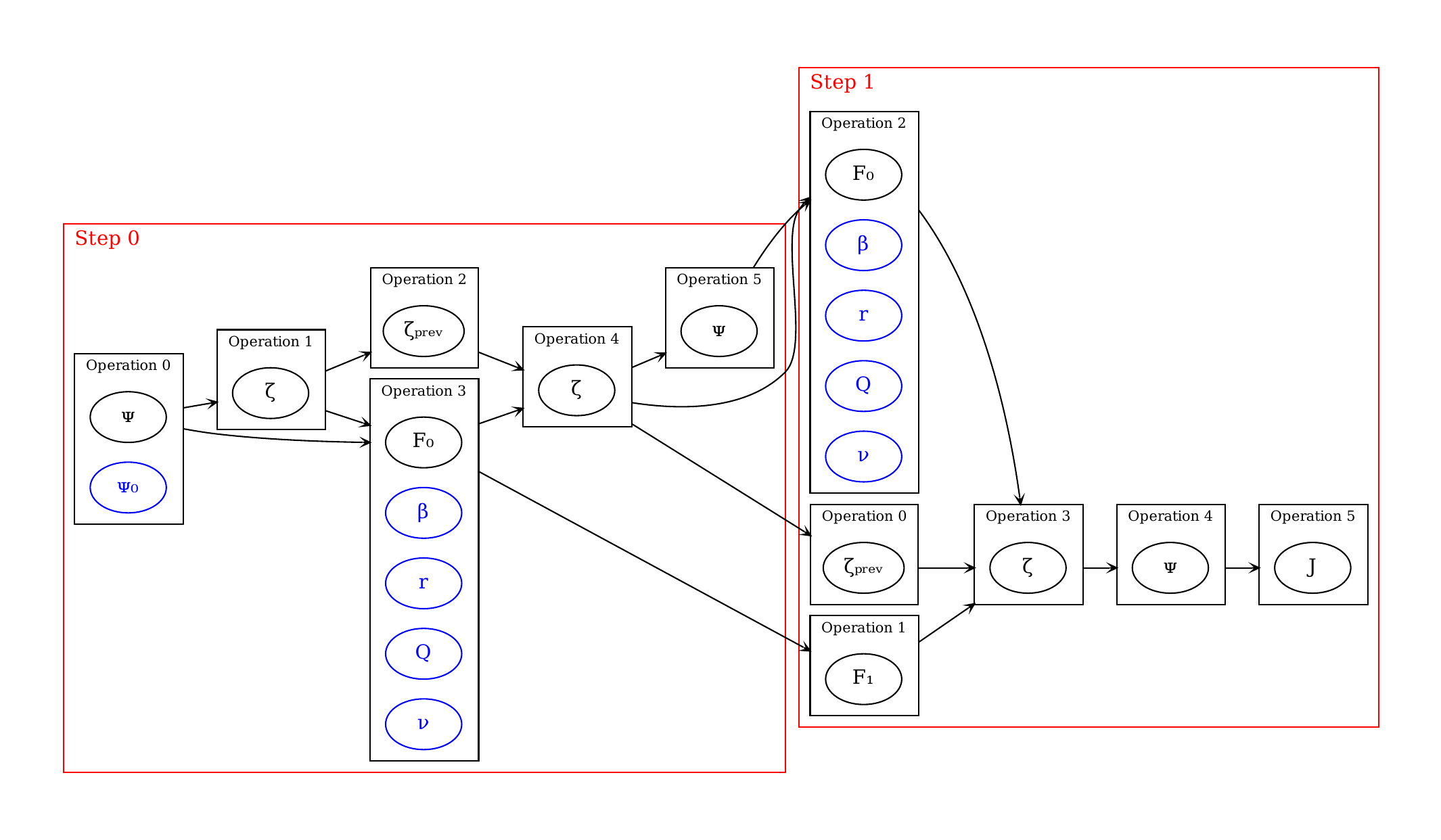}
\caption{Visualization of the computational graph for two timesteps in a solver for the barotropic vorticity equation. Step 0 corresponds to initialization and a forward Euler step, and step 1 to a second order Adams-Bashforth step and evaluation of a functional.}\label{fig:stommel_2}
\end{centering}\end{figure}

The computational graph for two timesteps of the numerical model is visualized in Figure~\ref{fig:stommel_2}. In this visualization the nodes of the graph correspond to the operations, computing values for the variables indicated in black. Directed edges indicate the earlier variables -- in terms of the operations that compute their values -- which are dependencies for later operations. Parameters are defined to be dependencies which do not have a value computed by earlier operations. In the visualization these are indicated in blue. The parameters could alternatively be introduced using additional nodes in the graph -- later an auxiliary step indicating parameters of interest will play a similar role. Steps are indicated with red rectangles. 

For example, operation 0 in step 0 corresponds to the assignment $\psi \gets \psi_0$, where the initial stream function $\psi_0$ is a parameter. Operation 5 in step 0 corresponds to the solution of a discrete Poisson equation, inverting the relative vorticity $\zeta$ to obtain the stream function $\psi$. Step 0 corresponds to initialization and the first timestep, and step 1 to the second timestep and the evaluation of a functional.

\subsection{An adjoint calculation}

A complete adjoint calculation consists of first evaluating all forward operations and then, in reverse order, computing values for adjoint variables associated with each forward operation.

\begin{figure}\begin{centering}
\includegraphics[width=1.0\textwidth]{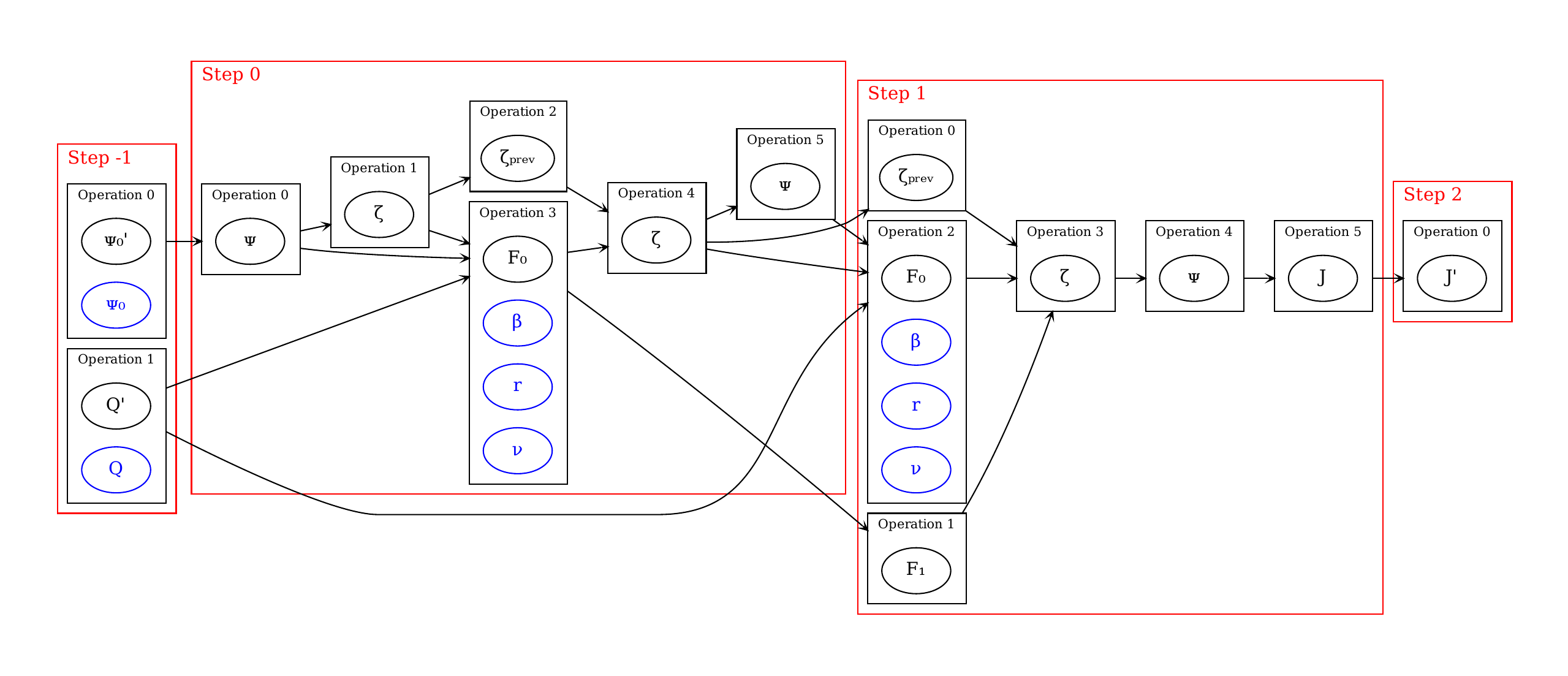}
\caption{As in Figure~\ref{fig:stommel_2}, but with the introduction of auxiliary steps to copy input parameters (the new step -1) and copy the output functional (the new step 2).}\label{fig:stommel_2_augmented}
\end{centering}\end{figure}

We first augment the forward with two additional steps. A first step, appearing at the start of the calculation, copies the values of input parameters of interest from a given input parameter to an auxiliary variable. A second step, appearing at the end of the calculation, copies the value of a functional of interest into an auxiliary output variable. The operations appearing in these auxiliary steps correspond to simple assignments, and their appearance simplifies the structure of the adjoint calculation to follow. An augmented model, considering the initial stream function $\psi_0$ and the wind forcing parameter $Q$ to be parameters of interest, is visualized in Figure~\ref{fig:stommel_2_augmented}.

Associated with each forward operation is a residual function which takes as input all dependencies and a candidate for the resulting output value. For example for operation 3 in step 0 in Figure~\ref{fig:stommel_2_augmented} we have a residual function $\mathcal{F}^0_3 \left( F_0, \beta, r, \nu, Q', \psi, \zeta \right)$ which takes as input the parameters $\beta$, $r$, and $\nu$, values for the forward variables $Q'$, $\psi$, and $\zeta$, and a candidate output value $F_0$. The output value is obtained by solving the root finding problem

\begin{equation*}
  \mathcal{F}^0_3 \left( \hat{F}_0, \beta, r, \nu, Q', \psi, \zeta \right) = 0,
\end{equation*}
to obtain $\hat{F}_0$, which is then assigned to the output variable. Since no ambiguity arises, the distinction between the candidate output value (here $F_0$) for a residual function and the output value obtained by solving the root finding problem is dropped.

The output for the $j$th operation in step $i$ is denoted $u^i_j$, with $i = -1$ corresponding to the auxiliary parameters step and $i = N$ corresponding to the auxiliary functional step. For simplicity it is assumed that the $u^i_j$ are real vectors, each with length $M^i_j$ for some positive integer $M^i_j$. Step $i$ consists of $N_i$ operations, with $N_i$ a positive integer, and the residual function for the $j$th operation in step $i$ is denoted $\mathcal{F}^i_j$ and has codomain $\mathbb{R}^{M^i_j}$. Residual functions in the auxiliary parameters and functional steps are defined such that their derivative with respect to the output is an identity matrix. Operations in each step are indexed in a forward causal sense -- in a forward calculation the calculation for $u^i_{j + 1}$ occurs after the calculation for $u^i_j$, and the calculation for $u^{i + 1}_0$ occurs after the calculation for $u^i_{N_i - 1}$. Associated with the $j$th operation in the $i$th step we introduce an adjoint variable $\lambda^i_j$ and an adjoint right-hand-side $b^i_j$, which are each real vectors with the same length as $u^i_j$. 

The adjoint calculation then proceeds according to Algorithm~\ref{alg:adjoint}. In Algorithm~\ref{alg:adjoint} the element in the $\alpha$th row and $\beta$th column of a matrix $\partial \mathcal{F}^i_j / \partial u^k_l$ contains the partial derivative of the $\alpha$th component of $F^i_j$ with respect to the $\beta$th component of $u^k_l$, each $\partial \mathcal{F}^i_j / \partial u^i_j$ is assumed invertible, and all vectors are column vectors. At the end of the calculation the adjoint variable $\lambda^{-1}_j$ is the derivative of the functional with respect to the parameter associated with the $j$th operation in the auxiliary parameters step.

\begin{algorithm}[h]
\KwResult{Sensitivities $\lambda^{-1}_j$}
\Begin{
\For{$i \gets -1$ to $N - 1$}{
  \For{$j \gets 0$ to $N_i - 1$}{
    $b^i_j \gets 0$\;
  }
}
$b^N_0 \gets 1$\;
\For{$i \gets N$ to $-1$}{
  \For{$j \gets N_i - 1$ to $0$}{
    Solve the adjoint linear system $\left( \partial \mathcal{F}^i_j / \partial u^i_j \right)^T \lambda^i_j = b^i_j$ for $\lambda^i_j$\;
    \For{each variable $u^k_l$ on which the calculation for $u^i_j$ depends}{
      $b^k_l \gets b^k_l - \left( \partial \mathcal{F}^i_j / \partial u^k_l \right)^T \lambda^i_j$\;
    }
  }
}
}
\caption{An adjoint calculation for a forward calculation whose variables are real vectors.}\label{alg:adjoint}
\end{algorithm}

An implementation of this algorithm can be optimized so that memory for a right-hand-side $b^i_j$ is allocated only when the first adjoint term is added, and to handle the (commonly encountered) case where $\partial \mathcal{F}^i_j / \partial u^i_j$ is an identity matrix. An activity analysis can be applied (e.g. \citet{griewank2008}, section 6.2) to avoid calculating adjoint terms or variables which do not depend implicitly on the adjoint initial condition $b^N_0$, and which do not implicitly influence the $\lambda^{-1}_j$ -- that is, $\lambda^i_j$ and terms contributing to $b^i_j$ need only be computed if, in the computational graph, $J$ is reachable from $u^i_j$ and $u^i_j$ is reachable from any $u^{-1}_k$.

A subset of the parameters and forward variables is now identified, consisting of only those whose values are needed to compute the matrices $\partial \mathcal{F}^i_j / \partial u^i_j$ and $\partial \mathcal{F}^i_j / \partial u^k_l$ which appear in Algorithm~\ref{alg:adjoint}  -- see also the definition of the \texttt{adjU} sets of variables identified in \citet{hascoet2005}. These are here referred to as the ``non-linear dependencies''.

The adjoint calculations associated with the auxiliary functional step with index $N$ consist of an assignment $\lambda^N_0 \gets b_0^N$ and an addition to one element of $b_l^k$, $b_{l,\alpha}^k \gets b_{l,\alpha}^k + \lambda^N_0$ for some $\alpha$, after which $b_{l,\alpha}^k = 1$. The auxiliary functional step facilitates initialization of the adjoint -- in practice this may simplify implementation in code, as code used to process the computational graph can be used to assist in initialization of the adjoint. The adjoint calculations associated with the auxiliary functional step with index $N$ are therefore extremely simple. The adjoint calculations associated with the auxiliary parameters step with index $-1$ are simple assignments, $\lambda^{-1}_j \gets b^{-1}_j$. Again, in practice the introduction of this step may simplify implementation in code, as the calculation of sensitivities need not be considered separately from other elements of the adjoint calculation. Since the cost of calculations in the auxiliary steps is expected to be negligible, checkpointing schedules need consider adjoint advances only over the $N$ steps with indices $N - 1$ to $0$ inclusive. For the remainder of this article we therefore do not explicitly include the auxiliary parameter and functional steps, but note that these can be added and used to facilitate the initialization of an adjoint calculation or the calculation of a sensitivity.

\section{Defining a checkpointing schedule}\label{sect:schedule}

A checkpointing schedule prescribes the combination of an original forward calculation together with at least one adjoint calculation, prescribing for example when checkpoints should be stored and loaded, and when the forward or adjoint should advance. In this section the requirements of a checkpointing schedule are discussed, incorporating a distinction between the data required to reinitialize and advance the forward, and data required to advance the adjoint.

Here a forward advance refers to the evaluation of operations in one or more steps, proceeding in a forward causal sense. An adjoint advance refers to the solutions of adjoint linear systems in one or more steps, together with the addition of contributions to adjoint right-hand-sides (possibly appearing in earlier steps), proceeding in a reverse causal sense. A ``checkpoint'' refers to data stored in some location (e.g. memory or disk) for later use by the schedule. The broad term ``checkpoint'' is used as checkpointing schedules which mix storage of forward restart and non-linear dependency data will later be considered. Storage of data sufficient to restart and advance the forward over a subsequent sequence of steps is here termed a ``forward restart checkpoint''.

\subsection{Forward and adjoint dependencies}

It is important to identify two classes of dependencies: the dependencies required to restart and advance the forward calculation, and the dependencies required to advance the adjoint calculation.

\begin{figure}\begin{centering}
\includegraphics[width=1.0\textwidth]{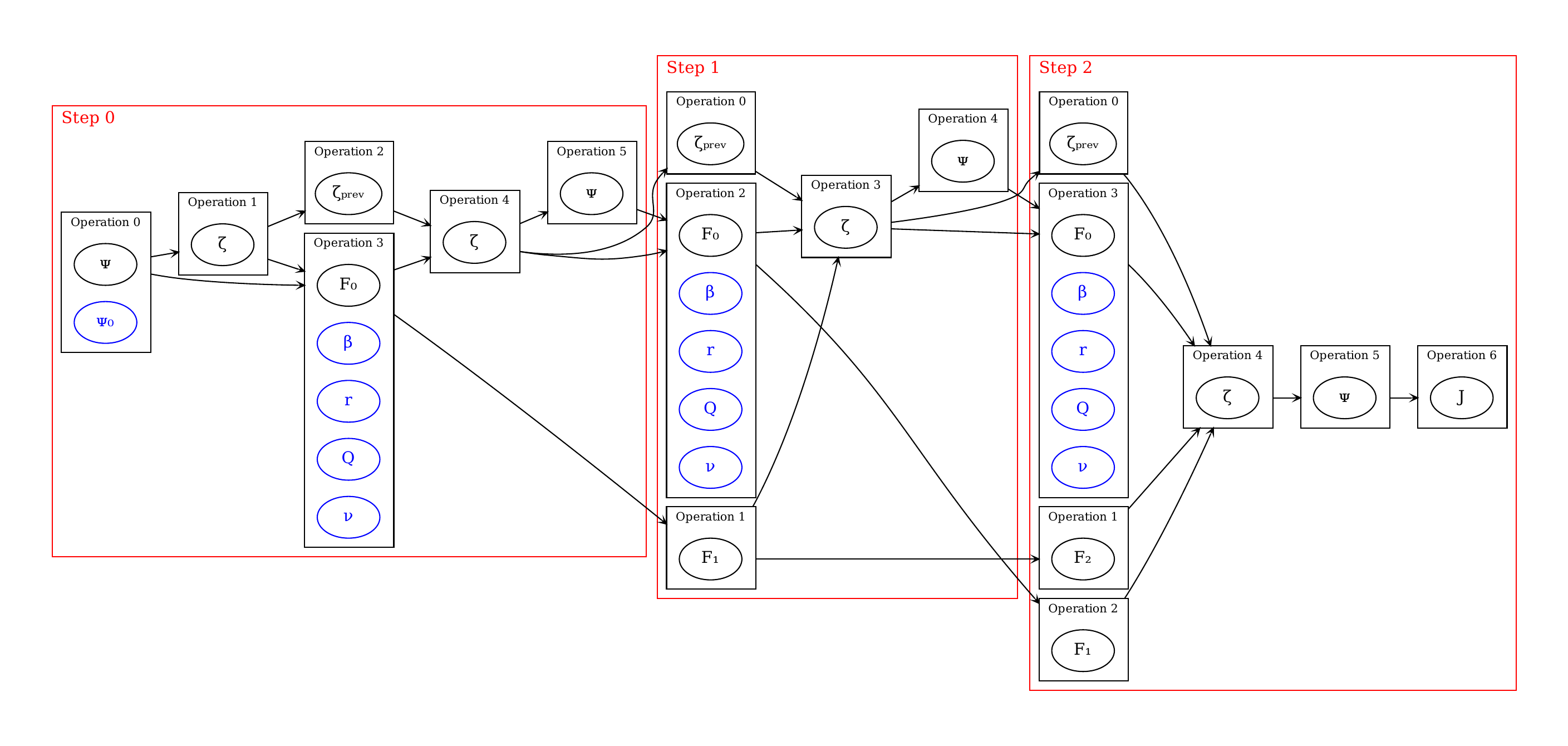}
\caption{Visualization of the computational graph for three timesteps in a solver for the barotropic vorticity equation. Step 0 corresponds to initialization and a forward Euler step, step 1 to a second order Adams-Bashforth step, and step 2 to a third order Adams-Bashforth step and evaluation of a functional.}\label{fig:dep_graph}
\end{centering}\end{figure}

Returning to the numerical solver for the barotropic vorticity equation, the computational graph associated with three timesteps is visualized in Figure~\ref{fig:dep_graph}. We can now consider, as an example, the dependencies which would need to be stored in a checkpoint associated with the start of step 1, sufficient for a forward advance over steps 1 and 2. These are given by the parameters $\beta$, $r$, $Q$, and $\nu$, as well as the variables $F_0$, $\zeta$, and $\psi$, computed respectively by operations 3, 4, and 5 in step 0.

We next consider the dependencies sufficient to advance the adjoint calculation over steps 2 and 1 -- that is, to solve for adjoint variables associated with each operation, and to add contributions to the adjoint right-hand-sides $b^k_l$ as in Algorithm~\ref{alg:adjoint}.  Note that we may add terms to adjoint right-hand-sides associated with earlier steps -- here step 0. The non-linear dependencies of the forward are sufficient for an adjoint calculation. For this example, ignoring possible non-linear dependencies associated with the functional $J$, for an adjoint advance over step 2 it suffices that the parameters $\beta$, $r$, and $\nu$, are stored, together with  $\zeta$ and $\psi$ computed respectively by operations 3 and 4 in step 1. For an adjoint advance over step 1 it suffices that the parameters $\beta$, $r$, and $\nu$, are stored, together with  $\zeta$ and $\psi$ computed respectively by operations 4 and 5 in step 0.

We can now consider making a choice between storing a forward restart checkpoint, sufficient to restart the forward at the start of step 1 and advance over steps 1 and 2, versus storing the non-linear dependencies for steps 1 and 2. In the case of a forward restart checkpoint, after loading the checkpoint the forward must in general advance to recompute non-linear dependency data. In the case where non-linear dependency data is stored no additional forward advance is needed. For example finite element assembly, required by the adjoint, can be performed directly if non-linear dependency data is available -- see Listing~\ref{lst:adjoint_assembly}. If parameters are ignored (which do not change throughout the calculation and might typically be stored separately in memory), and ignoring possible non-linear dependencies associated with the functional $J$, the forward restart checkpoint requires storage of three variables, while storing the non-linear dependencies directly requires storage of only two variables per adjoint step.

\begin{lstlisting}[language=Python, caption={An example of the calculation of an adjoint right-hand-side term using Firedrake. Here \texttt{F\_ij} is a \texttt{Form} object, and is a Unified Form Language symbolic representation for a residual. The first line computes a symbolic representation for a derivative, and the second line uses this to construct a symbolic representation for an adjoint right-hand-side term. The third line computes the value of the adjoint right-hand-side term via finite element assembly and adds it to an adjoint right-hand-side -- see Algorithm~\ref{alg:adjoint}. Only values for the forward dependencies appearing in the symbolic representation are needed for the finite element assembly. A high-level algorithmic differentiation tool can perform such calculations automatically.}, captionpos=b, label={lst:adjoint_assembly}, float, abovecaptionskip=1em]
    dF_ij_du_kl = derivative(F_ij, u_kl)
    adj_term = -action(adjoint(dF_ij_du_kl), lam_ij)
    b_kl += assemble(adj_term)
\end{lstlisting}

Note that, in general, the data which needs to be stored in a checkpoint to restart the forward calculation is dependent not only upon where the forward is to be restarted, but also to where it is to be advanced -- since later steps can depend on additional parameters or variables.

Note also that, in general, the non-linear dependencies required to advance the adjoint over a step need not be a subset of forward restart dependencies required to restart the forward at the start of that step. In the considered example, moving the calculations for the stream function from the end of one step to the start of the next step (e.g. moving operation 5 in step 0 to the start of step 1) would mean that these are no longer forward restart dependencies, but they would still be non-linear dependencies.

In \citet{griewank2000} it is indicated that steps should be chosen so that the amount of data required to advance the adjoint is at least as large as the amount of data stored in a checkpoint. To some extent these requirements can be met through a redefinition of the steps -- for example one might choose to include operations corresponding to more than one timestep in a step. A high-level view of the forward, or a static analysis of the forward, might perhaps facilitate such a definition of the steps.

A high-level approach might typically be expected to reduce the size of the adjoint dependency data. For example, when the forward is viewed in terms of high-level operations, complete solvers for time-dependent partial differential equations can be written and differentiated without the need for the algorithmic differentiation tool to build a record of procedure calls. Any intermediate variables involved in the lower-level calculations are also invisible to the high-level algorithmic differentiation tool, and so cannot generate further data dependencies. This observation is later used to motivate a checkpointing schedule which is optimal, in terms of the total number of forward steps, if the size of forward restart data and single-step non-linear dependency data is the same.

\subsection{Choosing the data to store in a checkpoint}

The revolve algorithm \citep{griewank2000} is optimal, in the sense that it requires a minimal number of forward steps to solve the forward and adjoint problems and minimizes the number of times a checkpoint is stored. Optimality of the number of forward steps is dependent upon the assumption that the forward must always advance after loading data from a checkpoint. H-Revolve \citep{herrmann2020} provides more advanced schedules, balancing computation and storage costs, but still defines the forward problem in terms of a chain involving the full forward solution on each step.

If checkpoints include non-linear dependency data, then the forward may not always need to advance after loading from a checkpoint. This is the approach used in the context of multi-stage Runge-Kutta schemes in \citet{zhang2021} and \citet{zhang2023}.

Alternatively, instead of storing the data required to restart a forward calculation at the start of a step, sufficient to advance the forward over a consecutive sequence of steps, one may instead store some or all of the forward variables computed \emph{within} those steps -- which may be advantageous if the calculation of some forward variables is expensive. This is potentially distinct from the storage of non-linear dependency data, but requires an additional balancing between storage and recomputation costs.

When storing data for later use in an adjoint calculation there is therefore a choice as to whether to store dependencies required to initialize the forward calculation, to store variables computed within the forward calculation, or to store dependencies required by the adjoint -- these may all differ and may overlap. A fully optimized strategy may need to apply some combination of approaches, balancing storage and recalculation costs. This is then further complicated by noting that the data to be stored in a forward restart checkpoint depends in general on the set of steps to which it applies, and not only on the index of the first step.

This article later seeks to improve performance by using the distinction between dependencies required to initialize the forward calculation, and dependencies required by the adjoint.

\subsection{Controlling intermediate storage via the schedule}

In the operator overloading approach to algorithmic differentiation applied in \citet{farrell2013}, the record of forward operations is constructed dynamically at runtime. This means for example that, in the initial forward calculation, the data which must be stored for a forward restart checkpoint is known only \emph{after} the forward has already advanced. This issue is addressed in \citet{maddison2019} by deferring storage of the checkpoint, by first buffering checkpoint data. Here a checkpointing schedule structure is defined which makes this buffering explicit.

An intermediate storage is introduced. The intermediate storage is used both to buffer forward data, assembling the data required for a checkpoint, and also to store non-linear dependency data for use by an adjoint calculation. The state of the intermediate storage is controlled via actions in the checkpointing schedule which enable or disable the storage of forward restart data, enable or disable storage of non-linear dependency data, and clear the intermediate storage.

The revolve algorithm assumes that additional storage -- beyond that allocated for checkpoints -- is available to store the dependencies of the adjoint necessary to advance the adjoint one step. In \citet{griewank2000} it is indicated that steps should be defined so that the amount of data required to advance the adjoint one step is at least as large as the amount of data stored in a checkpoint. With this assumption, and for checkpointing schedules that do not store both forward restart data and non-linear dependency data at the same time (which includes revolve schedules), the additional storage can be reused to buffer data for forward restart checkpointing. This additional storage is here referred to as the ``intermediate storage''.

The schedule consists of actions which control the storage and deletion of checkpoints, forward and adjoint advances, and also the state of the intermediate storage. Specifically a schedule consists of the following actions and parameters.
\begin{itemize}
  \item \texttt{Clear(clear\_ics, clear\_data)}: Clear the intermediate storage. If \linebreak \texttt{clear\_ics} is true, clear the buffer used to store forward restart data. If \texttt{clear\_data} is true, clear storage of non-linear dependency data. 
  \item \texttt{Configure(store\_ics, store\_data)}: Configure the intermediate storage. If \texttt{store\_ics} is true, enable buffering of forward restart data. If \texttt{store\_data} is true, enable storage of non-linear dependency data.
  \item \texttt{Write(n, storage)}: Transfer the intermediate storage to a checkpoint stored in the storage indicated by \texttt{storage}, yielding a checkpoint associated with step \texttt{n}.
  \item \texttt{Forward(n\_0, n\_1)}: Advance the forward from the start of step \texttt{n\_0} to the start of step \texttt{n\_1}.
  \item \texttt{Read(n, storage, delete)}. Load a checkpoint associated with step \texttt{n} from the storage indicated by \texttt{storage}, and store in the intermediate storage. If \texttt{delete} is true then the checkpoint should be deleted.
  \item \texttt{Reverse(n\_1, n\_0)}. Advance the adjoint from the start of step \texttt{n\_1} to the start of step \texttt{n\_0} (i.e. over steps \texttt{n\_1} - 1 to \texttt{n\_0} inclusive).
  \item \texttt{EndForward()}: Indicates that the original forward calculation has concluded.
  \item \texttt{EndReverse(exhausted)}: Indicates that an adjoint calculation has concluded. If \texttt{exhausted} is true then no further adjoint calculation is possible without a complete recalculation of the forward, and the schedule concludes. Otherwise further actions can be supplied for additional adjoint calculations.
\end{itemize}

\texttt{Read} actions direct that the loaded checkpoint data should be stored in the intermediate storage. After loading forward restart data, the forward calculation itself can be initialized from the intermediate storage. In addition, the following construction
\begin{itemize}
  \item \texttt{Read(2, RAM, True)}
  \item \texttt{Write(2, disk)}
\end{itemize}
can be used to transfer a checkpoint from memory to disk. Such transfers can occur, for example, in H-Revolve schedules \citep{herrmann2020}.

As an example, a revolve \texttt{youturn} action \citep{griewank2000} advances the forward one step while storing non-linear dependency data, and then advances the adjoint one step. With the above actions, and assuming the intermediate storage is initially empty, this becomes
\begin{itemize}
  \item \texttt{Configure(False, True)}: Disable storage of forward restart data, enable storage of non-linear dependency data.
  \item \texttt{Forward(n, n + 1)}: Advance the forward over the step.
  \item \texttt{Reverse(n + 1, n)}: Advance the adjoint over the step.
  \item \texttt{Clear(True, True)}: Clear the intermediate storage.
\end{itemize}

\subsection{Indicating the total number of steps}

The generation of a schedule always requires one piece of information from the application code: the total number of steps. This can be supplied to schedules either when they are intialized or, for checkpointing schedules which support it, can be defined after the original forward calculation. In the following an auxiliary action (not part of the schedule, but instead provided by the application code) \texttt{Initialize(max\_n)} indicates, at the start of the forward calculation, that the number of steps is \texttt{max\_n}.

\subsection{Implementation}

Checkpointing schedules are implemented in \tlmadjoint{} using Python generators. This approach allows a schedule to indicate a checkpointing action to perform, and to hand back control to the application code, while also maintaining the current state of the scheduler. The logical flow of code defining the schedule itself is maintained, simplifying the implementation of new schedules.

The schedules can be applied in general to any model which can be differentiated with the \tlmadjoint{} algorithmic differentiation tool. No further modification of application code, beyond the specification of forward steps and the definition of the schedule, is needed. All checkpointing schedules can, moreover, be applied to higher order adjoint calculations. In the reverse-over-forward approach used by \tlmadjoint{} tangent-linear operations are derived and then processed as new forward operations \citep{maddison2019}. This allows, for example, more advanced checkpointing schedules to be applied in the calculation of Hessian actions.

The ``non-linear dependencies'' as considered here are defined to be the forward dependencies of the adjoint. \tlmadjoint{} substitutes these with dependencies of all derivatives of forward residuals, which defines a superset of the desired non-linear dependencies. This may in particular include excess parameters or variables when an activity analysis is applied.

\tlmadjoint{} includes an implementation of the mixed memory-disk approach of \citet{stumm2009}, the approach combining periodic and binomial checkpointing described in \citet{pringle2016} and in the supporting information for \citet{goldberg2020}, and also provides two-level mixed memory-disk schedules by interfacing with the H-Revolve library \citep{herrmann2020}.

\subsection{A revolve schedule}\label{sect:revolve_schedule}

The schedule in Table~\ref{tab:binomial} corresponds to a revolve schedule for the case of 4 steps and a maximum of 2 forward restart checkpoints. The intermediate storage has one of three states: storing only forward data for a forward restart checkpoint (e.g. actions 0--3), storing only non-linear dependency data for the adjoint (e.g. actions 18--21), and storing no data (actions 15--17).

Note that, while it may break the assumptions of the revolve algorithm, the schedule permits a complete adjoint calculation even with an arbitrary computational graph. For example action 4 indicates that data for a forward restart checkpoint should be stored in the intermediate storage. This occurs in action 5 as the forward advances. The data is transferred to a checkpoint in action 6, and then the intermediate storage is cleared in action 7. Even if calculations in later steps depend on additional forward variables, not recorded in the forward restart checkpoint, these need not appear in the checkpoint. In this example any additional forward variables required to advance over step 3 need not be stored in the checkpoint, as the forward only advances over step 3 once (in action 9). More generally data required to advance the forward over later steps may appear in later forward restart checkpoints. Through explicit control of both the intermediate storage and forward advances via the schedule, it is possible to store only data necessary to advance the forward over a specific sequence of steps -- implementing checkpoint deferment as used in \citep{maddison2019}.

There is, however, potential inefficiency during the adjoint calculation, as the range of steps over which the forward needs to advance from a forward restart checkpoint may reduce, either as the adjoint advances or as new forward restart checkpoints are created. For example the forward restart checkpoint created in actions 4--7 contains data sufficient to restart and advance the forward over steps 1 and 2. At action 20 the adjoint advances over step 2, and any additional data required to advance the forward over step 2 could then in principle be removed from the checkpoint.

\begin{table}\begin{centering}
\begin{tabular}{c|l|c|c}
index & \texttt{action(parameters)} & \begin{tabular}{c} forward \\ state \end{tabular} & \begin{tabular}{c} adjoint \\ state \end{tabular} \\
\midrule
- & \texttt{Initialize(4)} & - & - \\
0 & \texttt{Configure(True, False)} & - & - \\
1 & \texttt{Forward(0, 1)} & $0 \rightarrow 1$ & - \\
2 & \texttt{Write(0, disk)} & - & - \\
3 & \texttt{Clear(True, True)} & - & - \\
4 & \texttt{Configure(True, False)} & - & - \\
5 & \texttt{Forward(1, 3)} & $1 \rightarrow 3$ & - \\
6 & \texttt{Write(1, disk)} & - & - \\
7 & \texttt{Clear(True, True)} & - & - \\
8 & \texttt{Configure(False, True)} & - & - \\
9 & \texttt{Forward(3, 4)} & $3 \rightarrow 4$ & - \\
10 & \texttt{EndForward()} & - & - \\
11 & \texttt{Reverse(4, 3)} & - & $4 \rightarrow 3$ \\
12 & \texttt{Clear(True, True)} & - & - \\
\midrule
13 & \texttt{Read(1, disk, False)} & $\rightarrow 1$ & - \\
14 & \texttt{Clear(True, True)} & - & - \\
15 & \texttt{Configure(False, False)} & - & - \\
16 & \texttt{Forward(1, 2)} & $1 \rightarrow 2$ & - \\
17 & \texttt{Clear(True, True)} & - & - \\
18 & \texttt{Configure(False, True)} & - & - \\
19 & \texttt{Forward(2, 3)} & $2 \rightarrow 3$ & - \\
20 & \texttt{Reverse(3, 2)} & - & $3 \rightarrow 2$ \\
21 & \texttt{Clear(True, True)} & - & - \\
\midrule
22 & \texttt{Read(1, disk, True)} & $\rightarrow 1$ & - \\
23 & \texttt{Clear(True, True)} & - & - \\
24 & \texttt{Configure(False, True)} & - & - \\
25 & \texttt{Forward(1, 2)} & $1 \rightarrow 2$ & - \\
26 & \texttt{Reverse(2, 1)} & - & $2 \rightarrow 1$ \\
27 & \texttt{Clear(True, True)} & - & - \\
\midrule
28 & \texttt{Read(0, disk, True)} & $\rightarrow 0$ & - \\
29 & \texttt{Clear(True, True)} & - & - \\
30 & \texttt{Configure(False, True)} & - & - \\
31 & \texttt{Forward(0, 1)} & $0 \rightarrow 1$ & - \\
32 & \texttt{Reverse(1, 0)} & - & $1 \rightarrow 0$ \\
33 & \texttt{Clear(True, True)} & - & - \\
34 & \texttt{EndReverse(True)} & - & -
\end{tabular}\caption{A revolve schedule converted to the described schedule structure. Here the original forward calculation consists of 4 steps, and there are a maximum of 2 permitted forward restart checkpoints at any one time. The forward and adjoint states refer to the start of the given steps, indexing from zero. The schedule is divided by adjoint advances, corresponding to the left panel of Figure~\ref{fig:schedule_4_2}. The forward advances 8 steps in total.}\label{tab:binomial}
\end{centering}\end{table}

\section{Mixing storage of forward restart and non-linear dependency data}\label{sect:sub_revolve}

Fully optimal schedules, where mixed storage of forward restart and non-linear dependency data are considered, require detailed knowledge of different costs. Here, instead, assumptions as used in the revolve algorithm are considered, with the additional assumption that forward restart and single-step non-linear dependency data sizes are the same. This demonstrates the potential of using more detailed knowledge of the computational graph for improved checkpointing performance, and demonstrates how a version of a CAMS-GEN schedule \citep{zhang2023} can be applied more generally for cases where the complete structure of the computational graph is determined dynamically at runtime.

\subsection{Assumptions}

It is assumed that the non-linear dependency data associated with any single step has the same size as the forward restart data associated with the start of any step. Other assumptions are as for the revolve algorithm. It is assumed that the number of forward steps in the original forward calculation is known at the start of the forward calculation. It is assumed that the forward restart checkpoint size is always the same, irrespective of the steps over which the forward is subsequently to be advanced (ignoring in particular additional costs associated with ``long-range'' dependencies). The performance of the schedule is defined in terms of the total number of forward steps taken -- which is a measure of runtime performance, excluding the cost of adjoint advances, if the time taken to solve each forward step is the same, and all runtime costs associated with storage are ignored. The forward and adjoint always advance over full steps, and (as in section 12.3 of \citet{griewank2008}) we do not permit the calculation of forward data for earlier steps using forward data for later steps. It is not assumed that non-linear dependency data for step $m$ suffices to restart the forward at the start of step $m + 1$.

Without increasing the total number of forward steps required, we can exclude the case where both forward restart data associated with the start of a step, and non-linear dependency data associated with the same step, are stored in checkpoints at the same time. This can be concluded by observing that we need only store forward restart data for a step $m$ in a checkpoint, allowing a restart at the start of step $m$, if we later wish to recompute forward data within that step. Rerunning the forward over step $m$ recomputes the non-linear dependency data for that step. Hence if the forward is later rerun over step $m$ we need not simultaneously store the non-linear dependency data in a checkpoint, and if the forward is not later rerun over step $m$ we need not store the forward restart data in a checkpoint. If a schedule results in both being stored in a checkpoint at the same time, we can therefore always modify the schedule so that only one is stored in a checkpoint at once, without increasing the total number of forward steps.

As in \citet{zhang2023} we consider $s$ ``checkpointing units'', which may be used to store forward restart or non-linear dependency data. Storing forward restart data, or non-linear dependency data for one step, each use one unit of storage, and are referred to respectively as a forward restart checkpoint or non-linear dependency data checkpoint. It is important to note, however, that while stored non-linear dependency data is sufficient to advance the adjoint over a step, it may not be sufficient to restart and advance the forward calculation. 

Given the observation above, we limit consideration to the case where a checkpoint associated with a step is used either to restart the forward at the start of the step, or to store non-linear dependency data for the step, but not both. We assume one additional unit of storage is available to store non-linear dependency data used to advance the adjoint one step, and use this as the intermediate storage.

The resulting schedule, mixing storage of forward restart data and non-linear dependency data in checkpoints, is referred to as a ``mixed schedule''.

\subsection{Dynamic programming problem}

Given $s$ remaining checkpointing units, we consider the problem of advancing the adjoint over $n$ steps -- i.e. advancing the adjoint to the start of step $n_0$, given that the forward is initially at the start of step $n_0$ and the adjoint is initially at the start of step $n_0 + n$.

\begin{figure}\begin{centering}
\begin{tabular}{c p{0.05\textwidth} c}
\includegraphics[width=0.4\textwidth,page=7,trim=50 30 50 30,clip]{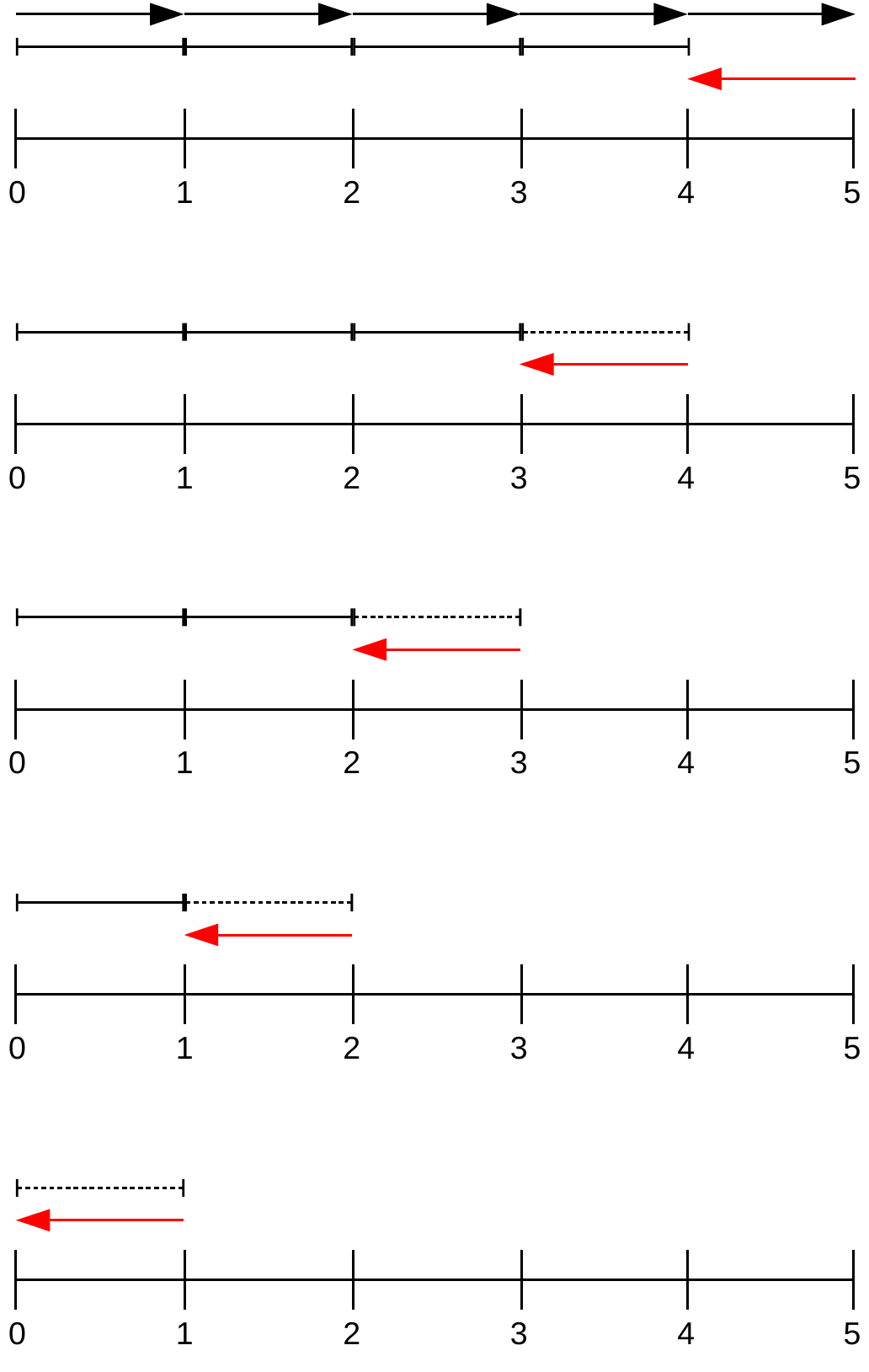}
& &
\includegraphics[width=0.4\textwidth,page=8,trim=50 30 50 30,clip]{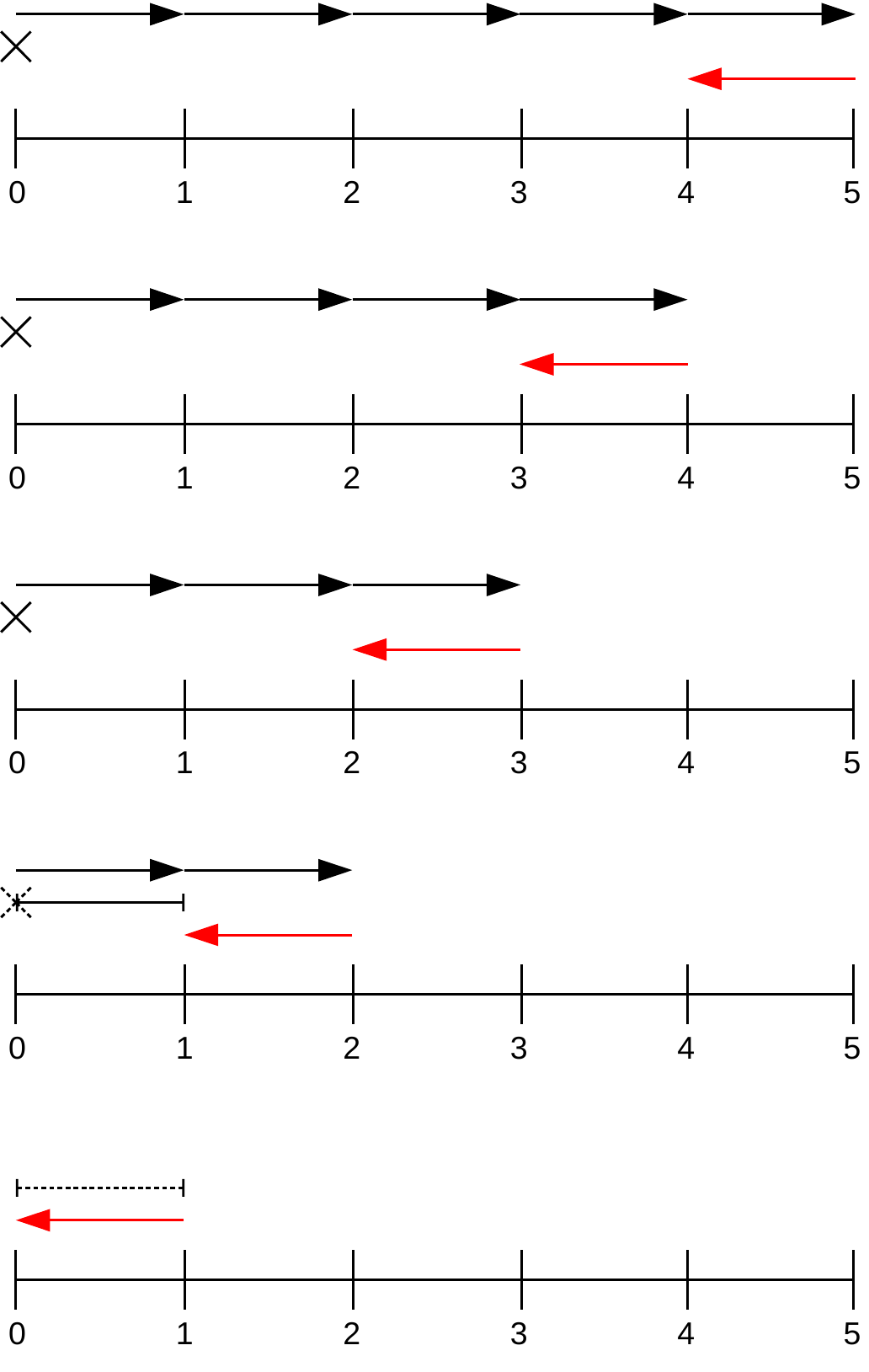} \\
Case 1, $n \le s + 1$
& &
Case 2, $s = 1$ and $n > 2$
\end{tabular}
\caption{Two options considered when constructing the schedule, illustrated for $n = 5$ steps. Left: Case 1, storage of all non-linear dependency data in checkpoints and the intermediate storage. Right: Case 2, a single checkpointing unit. The schedules proceed from top to bottom. The numbered labels indicate the start of a given step, counting from zero. The black arrows pointing to the right, at the top, indicate forward advances. Below this a filled cross indicates a forward restart checkpoint, and a filled line with end bars a non-linear dependency data checkpoint, with checkpoints either stored as part of the indicated forward advance, or retained from previous forward advances. Dashed versions of these indicate a checkpoint which is loaded and then deleted. Deletes occur before any new checkpoints are stored. Red arrows pointing to the left indicate adjoint advances, occurring after loading of checkpoints and forward advances.}\label{fig:dynamic_programming_1_2}
\end{centering}\end{figure}

\begin{figure}\begin{centering}
\begin{tabular}{c c}
\includegraphics[width=0.4\textwidth,page=9,trim=50 370 50 350,clip]{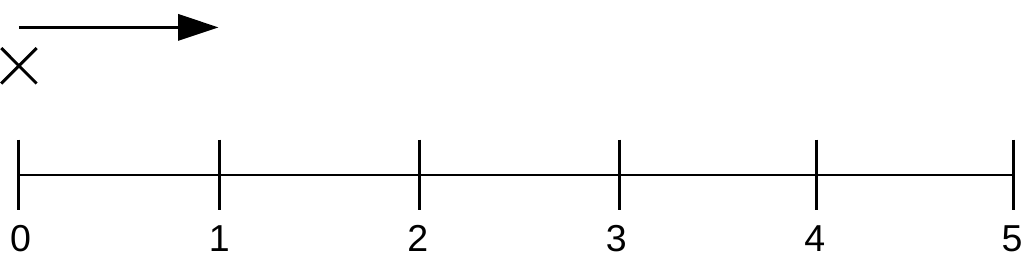} &
\includegraphics[width=0.4\textwidth,page=10,trim=50 370 50 350,clip]{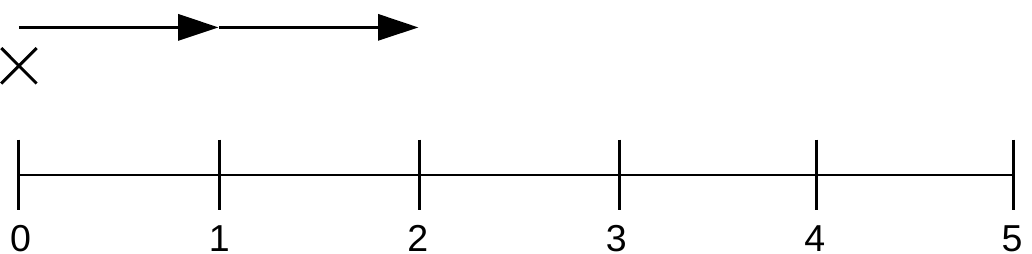} \\
Case 3 (a), $m = 1$ &
Case 3 (a), $m = 2$ \\
& \\
& \\
\includegraphics[width=0.4\textwidth,page=11,trim=50 370 50 350,clip]{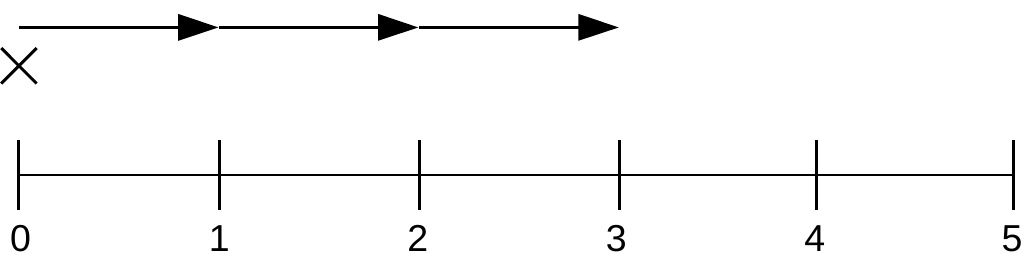} &
\includegraphics[width=0.4\textwidth,page=12,trim=50 370 50 350,clip]{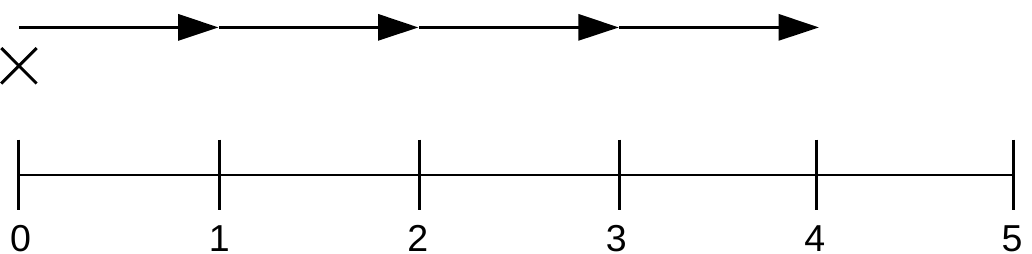} \\
& \\
Case 3 (a), $m = 3$ &
Case 3 (a), $m = 4$ \\
\end{tabular}
\caption{Options considered when constructing the schedule for case 3 (a), when $2 \le s \le n - 2$, illustrated for $n = 5$. For interpretation see Figure~\ref{fig:dynamic_programming_1_2}. Only the initial forward advance is shown. Storage of a forward restart checkpoint together with a forward advance of $m$ steps with $m \in \left\{ 1, \ldots, n - 1 \right\}$. There are $s - 1$ checkpointing units remaining for use when advancing the adjoint over the final $n - m$ steps. All $s$ checkpointing units can be used when advancing the adjoint over the first $m$ steps -- with the indicated forward restart checkpoint deleted, after it is loaded, if needed.}\label{fig:dynamic_programming_3a}
\end{centering}\end{figure}

\begin{figure}\begin{centering}
\begin{tabular}{c}
\includegraphics[width=0.4\textwidth,page=13,trim=50 370 50 350,clip]{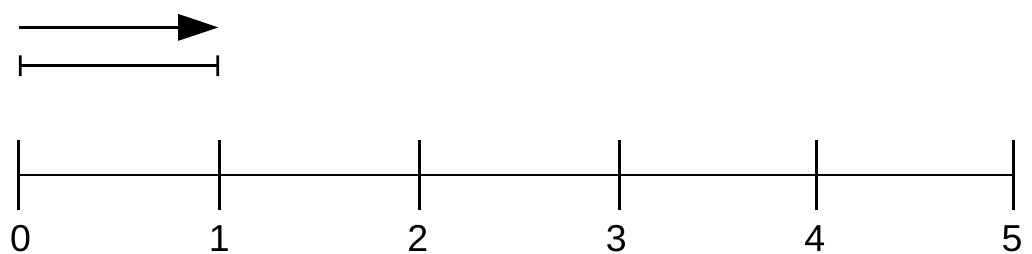} \\
Case 3 (b)
\end{tabular}
\caption{Option considered when constructing the schedule for case 3 (b), when $2 \le s \le n - 2$, illustrated for $n = 5$. For interpretation see Figure~\ref{fig:dynamic_programming_1_2}. Only the initial forward advance is shown. A forward advance of one step together with storage of a non-linear dependency data checkpoint. There are $s - 1$ checkpointing units remaining for use when advancing the adjoint over the final $n - 1$ steps. Data stored in the non-linear dependency data checkpoint can be used to advance the adjoint over the first step.}\label{fig:dynamic_programming_3b}
\end{centering}\end{figure}

A schedule is constructed by considering three cases.
\begin{enumerate}
  \item If $n \le s + 1$ then non-linear dependency data for the first $n - 1$ steps can be stored in checkpoints, and non-linear dependency data for the last step stored in the intermediate storage. $n$ forward steps are required. \label{item:case_minimal_rerun}
  \item If $s = 1$ and $n > 2$ then a forward restart checkpoint is stored at the start of the first step, and the forward is repeatedly advanced to recompute non-linear dependency data. When the adjoint has two steps left to advance over, the forward restart checkpoint is deleted and replaced with storage of non-linear dependency data for the first step, saving one forward step. $n \left( n + 1 \right) / 2 - 1$ forward steps are required. \label{item:case_maximal_rerun}
  \item Otherwise data is stored in a checkpointing unit, and the forward advances, using one of the following approaches.
  \begin{enumerate}
    \item Store a forward restart checkpoint associated with the start of the first step and advance $m$ steps, for some $m \in \left\{ 1, \ldots, n - 1 \right\}$. There are $s$ checkpointing units to advance the adjoint over the first $m$ steps, and $s - 1$ checkpointing units to advance the adjoint over the final $n - m$ steps. \label{item:case_forward_restart}
    \item Advance the forward one step, storing non-linear dependency data associated with the step in a checkpoint. There are $s - 1$ checkpointing units to advance the adjoint over the remaining $n - 1$ steps.
  \end{enumerate}
\end{enumerate}
Note that it is assumed that a forward restart checkpoint is not initially stored at the start of the first step. Checkpoints can be deleted when no longer needed, making checkpointing units available for other parts of the complete schedule. The different cases are illustrated in Figures~\ref{fig:dynamic_programming_1_2}, \ref{fig:dynamic_programming_3a}, and \ref{fig:dynamic_programming_3b}.

The minimal number of forward steps taken is defined by the dynamic programming problem\footnote{This notation differs from \citet{griewank2000} -- here the \emph{total} number of forward steps is considered.} (defined for positive integer $n$ and integer $s \ge \min \left( 1, n - 1 \right)$)
\begin{equation}\label{eqn:sub_revolve}
  p \left( n, s \right)
    = \begin{cases}
        n & \text{if } n \le s + 1, \\
        \frac{1}{2} n \left( n + 1 \right) - 1 &
          {\setlength{\arraycolsep}{0pt}
          \begin{array}{l} \text{if } s = 1 \\ \text{and } n > 2, \end{array}
          } \\
        \min \left\{ \begin{array}{l}
          \underset{m \in \left\{ 2, \ldots, n - 1 \right\}}{\min}
            \left[
          m + p \left( m, s \right)
          + p \left( n - m, s - 1 \right) \right] \\
          1 + p \left( n - 1, s - 1 \right)
          \end{array} \right. & \text{otherwise}.
      \end{cases}
\end{equation}
This has been simplified slightly in range of $m$ considered in the inner minimum, which follows as $p \left( m, s \right) > 0$ -- that is, the forward must always advance at least two steps after a forward restart checkpoint is stored.

\begin{figure}\begin{centering}
\includegraphics[width=0.45\textwidth]{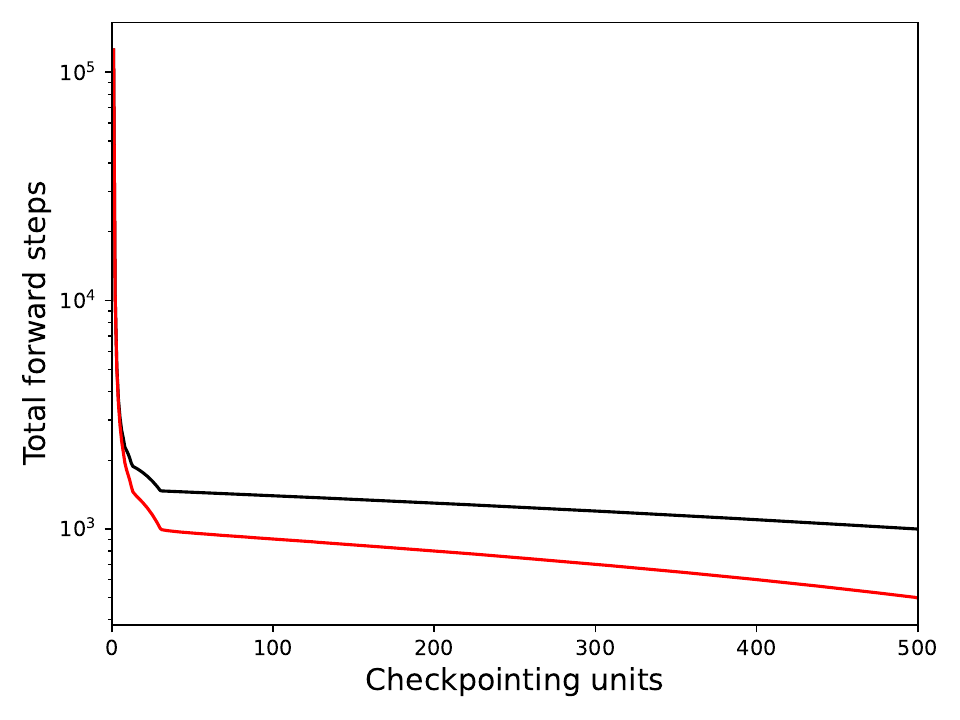}
~
\includegraphics[width=0.45\textwidth]{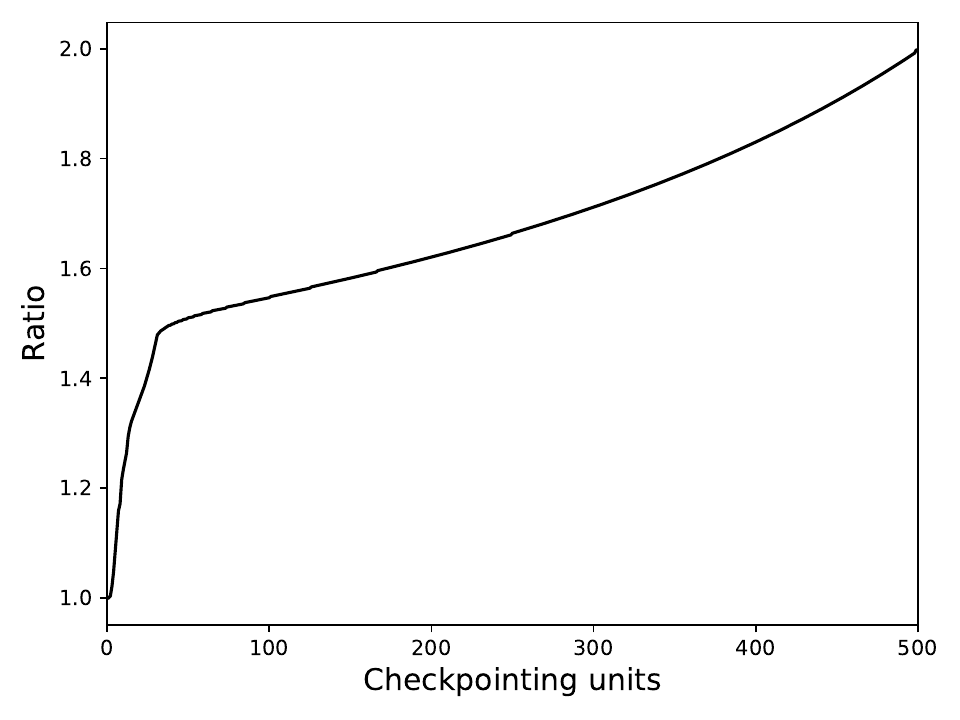}
\caption{Comparison of revolve versus a mixed schedule satisfying \eqref{eqn:sub_revolve}. Left: Total number of forward steps for a revolve schedule (black line) and a mixed schedule satisfying \eqref{eqn:sub_revolve} (red line). Right: The ratio: the total number of forward steps for the revolve schedule, divided by the total number of forward steps for the mixed schedule. $500$ forward steps in the original forward calculation are considered.}\label{fig:sub_revolve}
\end{centering}\end{figure}

Cases \ref{item:case_maximal_rerun} and \ref{item:case_forward_restart} are similar to the cases that appear in the dynamic programming problem associated with the revolve algorithm, except for the reduction by one step in the former. This can be seen in the similarity of elements of the dynamic programming problem \eqref{eqn:sub_revolve} to the dynamic programming problem appearing in \citet{griewank2000} (their equation (2)). The performance of a mixed schedule satisfying \eqref{eqn:sub_revolve} is demonstrated relative to the revolve algorithm in Figure~\ref{fig:sub_revolve}.

The solution to the dynamic programming problem \eqref{eqn:sub_revolve} is equivalent to the solution of the CAMS-GEN double dynamic programming problem for $l = 1$ stage \citep[][Lemmas 2 and 3 and Theorem 2]{zhang2023}, provided appropriate terminating cases corresponding to cases \ref{item:case_minimal_rerun} and \ref{item:case_maximal_rerun} above are used. However these terminating cases require the ability to later replace a forward restart checkpoint with a non-linear dependency data checkpoint.

\subsection{Constructing a schedule}

For the original forward calculation a schedule is constructed by solving the dynamic programming \eqref{eqn:sub_revolve}, keeping a record of the cases which lead to optimal solutions. Ties are broken by prioritizing storage of forward restart data over storage of non-linear dependency data, and by maximizing forward advancement when forward restart checkpoints are stored. In \tlmadjoint{} the dynamic programming problem \eqref{eqn:sub_revolve} is solved using tabulation, with the key section of code just-in-time compiled using Numba \citep{lam2015}. Memoization is used if Numba is not available. After the original forward calculation the adjoint can advance one step, using non-linear dependency data stored in the intermediate storage.

To continue the schedule, the most recent checkpoint is loaded. A new solution to the dynamic programming problem is used to identify whether the checkpoint should be deleted, with ties broken as before. For example if a forward restart checkpoint associated with the start of step $n_0'$ is loaded, with the adjoint at the start of step $n_0' + n'$, and there are $s'$ checkpointing units currently available, then the solution of the dynamic programming problem for $n = n'$ and $s = s' + 1$ determines which checkpoint associated with step $n_0'$ should be stored. If this indicates that a forward restart checkpoint should be stored, then the loaded forward restart checkpoint is retained, and otherwise it is deleted. It can inferred that a non-linear dependency data checkpoint can be deleted as soon as it is loaded. The new solution to the dynamic programming problem allows the adjoint to advance one more step, and the process then repeats for the complete adjoint calculation.

\begin{figure}\begin{centering}
\begin{tabular}{c p{0.05\textwidth} c}
\includegraphics[width=0.4\textwidth,page=3,trim=80 100 80 100,clip]{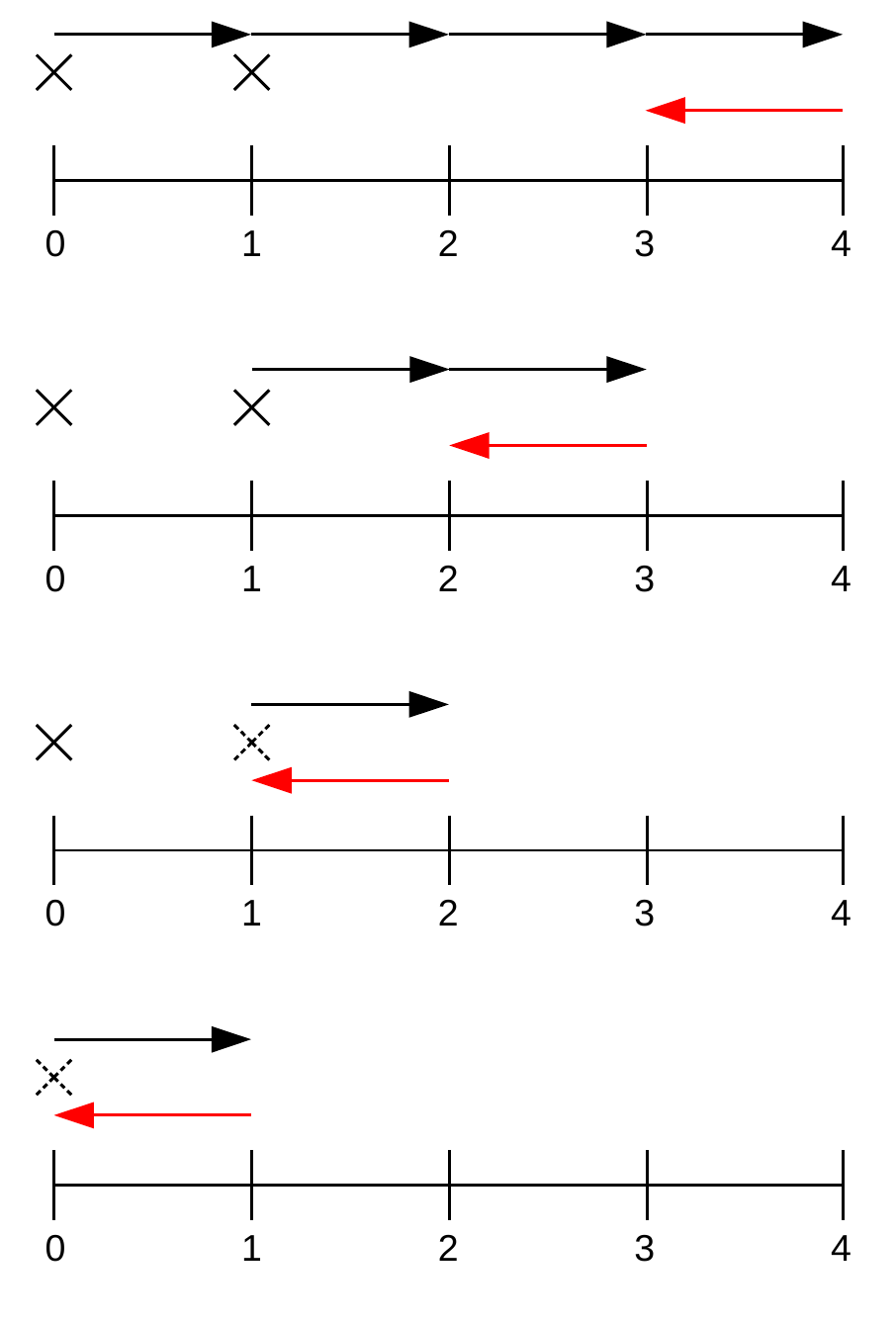}
& &
\includegraphics[width=0.4\textwidth,page=6,trim=80 100 80 100,clip]{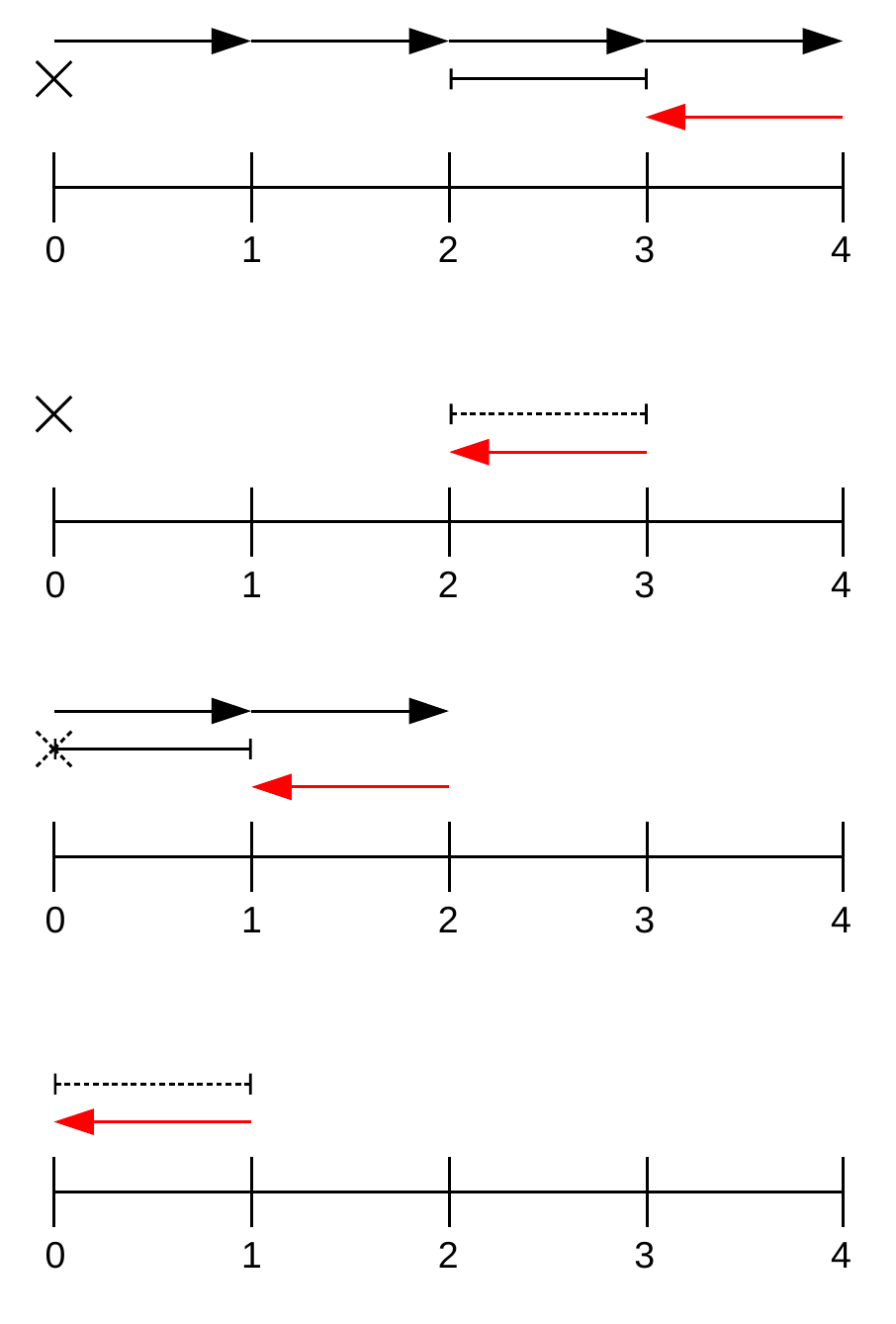} \\
Revolve & & Mixed
\end{tabular}
\caption{Schematics of checkpointing schedules for the case of 4 forward steps and 2 checkpointing units. For interpretation see Figure~\ref{fig:dynamic_programming_1_2}. Left: A revolve schedule, taking 8 forward steps in total. Right: A mixed schedule, taking 6 forward steps in total.}\label{fig:schedule_4_2}
\end{centering}\end{figure}

A resulting mixed schedule for the case of 4 forward steps and with 2 checkpointing units is shown schematically in the right panel of Figure~\ref{fig:schedule_4_2}, and the full schedule listed in Table~\ref{tab:sub_revolve}. Note that non-linear dependency data is stored in checkpoints using actions 4--7 and 18--21, and loaded in actions 13 and 26. Note also that a forward restart checkpoint is deleted in action 16, before a non-linear dependency data checkpoint is stored in actions 18--21.

\begin{table}\begin{centering}
\begin{tabular}{c|l|c|c}
index & \texttt{action(parameters)} & \begin{tabular}{c} forward \\ state \end{tabular} & \begin{tabular}{c} adjoint \\ state \end{tabular} \\
\midrule
- & \texttt{Initialize(4)} & - & - \\
0 & \texttt{Configure(True, False)} & - & - \\
1 & \texttt{Forward(0, 2)} & $0 \rightarrow 2$ & - \\
2 & \texttt{Write(0, disk)} & - & - \\
3 & \texttt{Clear(True, True)} & - & - \\
4 & \texttt{Configure(False, True)} & - & - \\
5 & \texttt{Forward(2, 3)} & $2 \rightarrow 3$ & - \\
6 & \texttt{Write(2, disk)} & - & - \\
7 & \texttt{Clear(True, True)} & - & - \\
8 & \texttt{Configure(False, True)} & - & - \\
9 & \texttt{Forward(3, 4)} & $3 \rightarrow 4$ & - \\
10 & \texttt{EndForward()} & - & - \\
11 & \texttt{Reverse(4, 3)} & - & $4 \rightarrow 3$ \\
12 & \texttt{Clear(True, True)} & - & - \\
\midrule
13 & \texttt{Read(2, disk, True)} & $\rightarrow \star$ & - \\
14 & \texttt{Reverse(3, 2)} & - & $3 \rightarrow 2$ \\
15 & \texttt{Clear(True, True)} & - & - \\
\midrule
16 & \texttt{Read(0, disk, True)} & $\rightarrow 0$ & - \\
17 & \texttt{Clear(True, True)} & - & - \\
18 & \texttt{Configure(False, True)} & - & - \\
19 & \texttt{Forward(0, 1)} & $0 \rightarrow 1$ & - \\
20 & \texttt{Write(0, disk)} & - & - \\
21 & \texttt{Clear(True, True)} & - & - \\
22 & \texttt{Configure(False, True)} & - & - \\
23 & \texttt{Forward(1, 2)} & $1 \rightarrow 2$ & - \\
24 & \texttt{Reverse(2, 1)} & - & $2 \rightarrow 1$ \\
25 & \texttt{Clear(True, True)} & - & - \\
\midrule
26 & \texttt{Read(0, disk, True)} & $\rightarrow \star$ & - \\
27 & \texttt{Reverse(1, 0)} & - & $1 \rightarrow 0$ \\
28 & \texttt{Clear(True, True)} & - & - \\
29 & \texttt{EndReverse(True)} & - & -
\end{tabular}\caption{A mixed checkpointing schedule. The forward consists of 4 steps, and there are 2 checkpointing units. A change in forward state $\rightarrow *$ indicates that, in general, the data loaded from a checkpoint is insufficient to restart the forward. The forward advances 6 steps in total. The schedule is divided by adjoint advances, corresponding to the right panel of Figure~\ref{fig:schedule_4_2}.}\label{tab:sub_revolve}
\end{centering}\end{table}

While it may break the data size assumptions, the checkpointing schedule described in this section permits a complete adjoint calculation even with an arbitrary computational graph. The structure of the computational graph can be determined dynamically at runtime, and no specific structure need be assumed for a valid adjoint calculation.

\subsection{Numerical example}\label{sect:benchmark}

In the above discussion the performance of a checkpointing schedule was measured in terms of the total number of forward steps. A numerical example is now considered to demonstrate that improved performance is achievable in practice, comparing a mixed schedule against a revolve schedule, with the schedules implemented in the \tlmadjoint{} library. The mixed schedule is applied to a solver for a time dependent partial differential equation making use of a linear multistep discretization for the time dimension. This example demonstrates the ability to apply the schedule even when the structure of the computational graph is determined dynamically at runtime, and tests a practical case where the forward calculation does not precisely align with the assumptions used to define the schedule.

The barotropic vorticity equation example is again considered, but is now integrated for a larger number of steps. Specifically we consider an implementation using Firedrake and \tlmadjoint{}, using versions of software as in \citep{firedrake_a,tlm_adjoint_c,stommel_benchmark_b}. Parameters as in \citet{stommel1948} are used, with a reduced drag coefficient, with the non-linear advection term retained, and with a Laplacian viscosity term. The domain is divided into a structured uniform triangle mesh formed by subdivision of a $256 \times 256$ grid of quadrilaterals into two triangles each. $17520$ timesteps are considered, with $200$ checkpointing units and with checkpoints stored in memory. Linear systems are solved by Cholesky factorization using PETSc \citep{petsc-user-3.20}. The reverse mode calculation computes the action of a Hessian on a single direction, with the Hessian defined by differentiating the time integrated kinetic energy twice with respect to the wind forcing term appearing on the right-hand-side of the vorticity equation. The Hessian action calculation is performed using a reverse-over-forward approach as described in \citet{maddison2019}, meaning that the full ``forward'' calculation recorded by \tlmadjoint{} includes tangent-linear operations as well as forward operations. Parameters, such as the initial stream function and tangent-linear direction, are excluded from checkpoints. Checkpointing schedule steps are defined to coincide with model timesteps.

\tlmadjoint{} caches finite element assembly data and linear solver data using an approach based on \citet{maddison2014}. Cached data can be shared between forward, tangent-linear, adjoint, and higher order adjoint calculations. \tlmadjoint{} further applies an activity analysis in both the calculation of adjoint variables, and also when advancing the forward during the reverse mode calculation.

The construction of revolve schedules can be simplified by noting that the associated dynamic programming problem involves a convex function \citep{griewank2000}. Here $p \left( n, s \right)$ defined in \eqref{eqn:sub_revolve} is non-convex in $n$, and hence no such simplification is apparent, and the generation of the mixed restart/non-linear dependency schedules has a significant cost for a problem of this size.

The runtime performance is tested in serial on a system with an Intel Core i5-10310U processor. The runtime of different elements is measured using the Python \texttt{time.perf\_counter} function. Performance is measured after an initial run, and the mean of three subsequent runs is measured. The Python ``oldest generation'' heuristic is disabled by manually patching CPython so that full collections occur even if there are a large number of long-lived objects,\footnote{See \url{https://devguide.python.org/internals/garbage-collector}, accessed 2024-04-30.} and garbage collection thresholds are doubled from their default values. The calculation using the revolve schedule takes $52358$ forward steps in total, and the calculation using the mixed schedule takes $34965$ forward steps in total. Mean runtimes are given in Table~\ref{tab:benchmark_runtime}. The mixed schedule leads to a significantly reduced reverse mode runtime, and the performance gain is larger than the additional cost of solving \eqref{eqn:sub_revolve}.

\begin{figure}\begin{centering}
\includegraphics[width=0.8\textwidth]{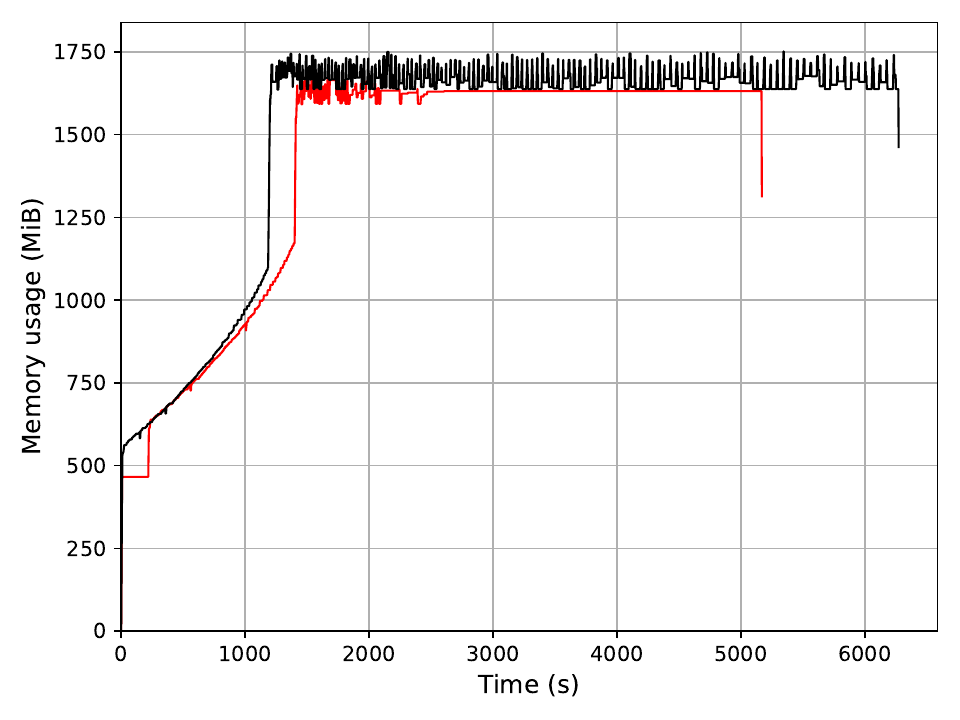}
\caption{Resident Set Size recorded for a complete barotropic vorticity calculation, using a revolve schedule (black) and a mixed schedule satisfying \eqref{eqn:sub_revolve} (red).}\label{fig:benchmark_memory}
\end{centering}\end{figure}

\tlmadjoint{} builds an unrolled record of operations in the time loop. While some optimizations are applied -- in particular the objects representing operations can be reused across different steps -- this means that there are parts of the calculation which have a storage cost linear in the number of steps. To test the memory usage, a further run of each case is conducted measuring the Resident Set Size throughout the calculation. This is shown in Figure~\ref{fig:benchmark_memory}. The revolve and mixed schedules lead to comparable peak memory usage.

\begin{table}\begin{centering}
\begin{tabular}{c|c|c|c}
Schedule & Tabulation & Forward & Reverse \\
\midrule
revolve & -- & 1191.1 & 5091.7 \\
mixed & 214.2 & 1189.1 & 3762.0
\end{tabular}\caption{Mean runtimes, in seconds, for a revolve schedule versus a mixed schedule. The tabulation time is the time taken to solve \eqref{eqn:sub_revolve}, with the solution optimized using Numba. Other initialization costs, including the initialization of the mesh and discrete function spaces, are not included in the reported runtimes. The ``forward'' time is the time taken for the combined forward and tangent-linear calculation. The ``reverse'' time is the time taken to compute a Hessian action, which includes solution of first and second order adjoint problems.}\label{tab:benchmark_runtime}
\end{centering}\end{table}

For this problem (and excluding the first two steps) a forward restart checkpoint stores data for $8$ finite element functions (together with the functional and its tangent-linear), while a non-linear dependency data checkpoint stores data for $6$ finite element functions. Although not applied here, the size of each could be reduced by $2$ finite element functions if the operations computing the stream function and integrating the kinetic energy (and their associated tangent-linears) were moved from the end of one step to the start of the next. It is possible, in either case, that performance might be improved by using more precise knowledge of the relative sizes of the two checkpoint types -- for example via a generalization of the CAMS-GEN scheme for non-integer $l < 1$.

\section{Conclusions}\label{sect:conclusions}

A checkpointing schedule structure has been introduced which explicitly controls an intermediate storage ``buffer'', and which can be used to defer storage of checkpoints as the forward calculation progresses, and the record of forward operations is constructed. This allows step-based checkpointing schedules to be used even when the record of forward operations is constructed dynamically at runtime.

This article has focused on how the simplified structure which arises with a high-level algorithmic differentiation approach -- appearing in the form of a simplified computational graph -- can be used when applying checkpointing strategies.  With the high-level approach it is possible to distinguish between the data required to restart and advance the forward, and the data required to advance the adjoint. This allows the use of checkpointing strategies which mix storage of forward restart and non-linear dependency data in checkpoints.

An optimal schedule is defined only for a given performance model. In practice the relative costs of different elements of the calculation will depend on details of the implementation. Runtime performance, and storage performance and limits, will also depend on the details of the system on which the calculation is performed. In the context of automated code generation these details appear below the level of the domain specific language meaning that, for a separation between application development and implementation optimization to be maintained, the development of higher performance checkpointing approaches itself requires automation.

\subsection*{Data availability}

\tlmadjoint{} is available at https://github.com/tlm-adjoint/tlm\_adjoint. The version as described in this article is available at \citep{tlm_adjoint_c}. Scripts for the benchmark in Section~\ref{sect:benchmark} are available at \citep{stommel_benchmark_b}.

\subsection*{Acknowledgements}

This work was supported by the Natural Environment Research Council [NE/T001607/1].

This research was funded in whole, or in part, by the Natural Environment Research Council [NE/T001607/1]. For the purpose of open access, the author has applied a creative commons attribution (CC BY) licence to any author accepted manuscript version arising.

JRM acknowledges useful communications with, and code contributions by, David A. Ham.

JRM would like to thank the three reviewers for their helpful comments.

\bibliographystyle{elsarticle-num-names}
\bibliography{references}

\begin{thebibliography}{37}
\expandafter\ifx\csname natexlab\endcsname\relax\def\natexlab#1{#1}\fi
\providecommand{\url}[1]{\texttt{#1}}
\providecommand{\href}[2]{#2}
\providecommand{\path}[1]{#1}
\providecommand{\DOIprefix}{doi:}
\providecommand{\ArXivprefix}{arXiv:}
\providecommand{\URLprefix}{URL: }
\providecommand{\Pubmedprefix}{pmid:}
\providecommand{\doi}[1]{\href{http://dx.doi.org/#1}{\path{#1}}}
\providecommand{\Pubmed}[1]{\href{pmid:#1}{\path{#1}}}
\providecommand{\bibinfo}[2]{#2}
\ifx\xfnm\relax \def\xfnm[#1]{\unskip,\space#1}\fi
\bibitem[{Farrell et~al.(2013)Farrell, Ham, Funke, and Rognes}]{farrell2013}
\bibinfo{author}{P.~E. Farrell}, \bibinfo{author}{D.~A. Ham},
  \bibinfo{author}{S.~W. Funke}, \bibinfo{author}{M.~E. Rognes},
\newblock \bibinfo{title}{Automated derivation of the adjoint of high-level
  transient finite element programs},
\newblock \bibinfo{journal}{SIAM Journal on Scientific Computing}
  \bibinfo{volume}{35} (\bibinfo{year}{2013}) \bibinfo{pages}{C369--C393}.
  \DOIprefix\doi{10.1137/120873558}.
\bibitem[{Aln\ae{}s et~al.(2014)Aln\ae{}s, Logg, \O{}lgaard, Rognes, and
  Wells}]{alnaes2014}
\bibinfo{author}{M.~S. Aln\ae{}s}, \bibinfo{author}{A.~Logg},
  \bibinfo{author}{K.~B. \O{}lgaard}, \bibinfo{author}{M.~E. Rognes},
  \bibinfo{author}{G.~N. Wells},
\newblock \bibinfo{title}{{U}nified {F}orm {L}anguage: A domain-specific
  language for weak formulations of partial differential equations},
\newblock \bibinfo{journal}{ACM Transactions on Mathematical Software}
  \bibinfo{volume}{40} (\bibinfo{year}{2014}) \bibinfo{pages}{9:1--9:37}.
  \DOIprefix\doi{10.1145/2566630}.
\bibitem[{Logg et~al.(2012)Logg, Mardal, and Wells}]{logg2012}
\bibinfo{editor}{A.~Logg}, \bibinfo{editor}{K.-A. Mardal},
  \bibinfo{editor}{G.~N. Wells} (Eds.), \bibinfo{title}{Automated solution of
  differential equations by the finite element method},
  volume~\bibinfo{volume}{84} of \textit{\bibinfo{series}{Lecture Notes in
  Computational Science and Engineering}}, \bibinfo{publisher}{Springer-Verlag
  Berlin Heidelberg}, \bibinfo{year}{2012}.
  \DOIprefix\doi{10.1007/978-3-642-23099-8}.
\bibitem[{Aln\ae{}s et~al.(2015)Aln\ae{}s, Blechta, Hake, Johansson, Kehlet,
  Logg, Richardson, Ring, Rognes, and Wells}]{alnaes2015}
\bibinfo{author}{M.~S. Aln\ae{}s}, \bibinfo{author}{J.~Blechta},
  \bibinfo{author}{J.~Hake}, \bibinfo{author}{A.~Johansson},
  \bibinfo{author}{B.~Kehlet}, \bibinfo{author}{A.~Logg},
  \bibinfo{author}{C.~Richardson}, \bibinfo{author}{J.~Ring},
  \bibinfo{author}{M.~E. Rognes}, \bibinfo{author}{G.~N. Wells},
\newblock \bibinfo{title}{The {FEniCS} project version 1.5},
\newblock \bibinfo{journal}{Archive of Numerical Software} \bibinfo{volume}{3}
  (\bibinfo{year}{2015}) \bibinfo{pages}{9--23}.
  \DOIprefix\doi{10.11588/ans.2015.100.20553}.
\bibitem[{Mitusch et~al.(2019)Mitusch, Funke, and Dokken}]{mitusch2019}
\bibinfo{author}{S.~K. Mitusch}, \bibinfo{author}{S.~W. Funke},
  \bibinfo{author}{J.~S. Dokken},
\newblock \bibinfo{title}{{d}olfin-adjoint 2018.1: automated adjoints for
  {FEniCS} and {F}iredrake},
\newblock \bibinfo{journal}{The Journal of Open Source Software}
  \bibinfo{volume}{4} (\bibinfo{year}{2019}).
  \DOIprefix\doi{10.21105/joss.01292}.
\bibitem[{Griewank and Walther(2008)}]{griewank2008}
\bibinfo{author}{A.~Griewank}, \bibinfo{author}{A.~Walther},
  \bibinfo{title}{Evaluating derivatives}, \bibinfo{edition}{second} ed.,
  \bibinfo{publisher}{Society for Industrial and Applied Mathematics},
  \bibinfo{year}{2008}.
\bibitem[{Baydin et~al.(2018)Baydin, Pearlmutter, Radul, and
  Siskind}]{baydin2018}
\bibinfo{author}{A.~G. Baydin}, \bibinfo{author}{B.~A. Pearlmutter},
  \bibinfo{author}{A.~A. Radul}, \bibinfo{author}{J.~M. Siskind},
\newblock \bibinfo{title}{Automatic differentiation in machine learning: a
  survey},
\newblock \bibinfo{journal}{Journal of Machine Learning Research}
  \bibinfo{volume}{18} (\bibinfo{year}{2018}) \bibinfo{pages}{1--43}.
\bibitem[{Maddison and Farrell(2014)}]{maddison2014}
\bibinfo{author}{J.~R. Maddison}, \bibinfo{author}{P.~E. Farrell},
\newblock \bibinfo{title}{Rapid development and adjoining of transient finite
  element models},
\newblock \bibinfo{journal}{Computer Methods in Applied Mechanics and
  Engineering} \bibinfo{volume}{276} (\bibinfo{year}{2014})
  \bibinfo{pages}{95--121}. \DOIprefix\doi{10.1016/j.cma.2014.03.010}.
\bibitem[{Griewank and Walther(2000)}]{griewank2000}
\bibinfo{author}{A.~Griewank}, \bibinfo{author}{A.~Walther},
\newblock \bibinfo{title}{Algorithm 799: Revolve: An implementation of
  checkpointing for the reverse or adjoint mode of computational
  differentiation},
\newblock \bibinfo{journal}{ACM Transactions on Mathematical Software}
  \bibinfo{volume}{26} (\bibinfo{year}{2000}) \bibinfo{pages}{19--45}.
  \DOIprefix\doi{10.1145/347837.347846}.
\bibitem[{Griewank(1992)}]{griewank1992}
\bibinfo{author}{A.~Griewank},
\newblock \bibinfo{title}{Achieving logarithmic growth of temporal and spatial
  complexity in reverse automatic differentiation},
\newblock \bibinfo{journal}{Optimization Methods and Software}
  \bibinfo{volume}{1} (\bibinfo{year}{1992}) \bibinfo{pages}{35--54}.
  \DOIprefix\doi{10.1080/10556789208805505}.
\bibitem[{Wang et~al.(2009)Wang, Moin, and Iaccarino}]{wang2009}
\bibinfo{author}{Q.~Wang}, \bibinfo{author}{P.~Moin},
  \bibinfo{author}{G.~Iaccarino},
\newblock \bibinfo{title}{Minimal repetition dynamic checkpointing algorithm
  for unsteady adjoint calculation},
\newblock \bibinfo{journal}{SIAM Journal on Scientific Computing}
  \bibinfo{volume}{31} (\bibinfo{year}{2009}) \bibinfo{pages}{2549--2567}.
  \DOIprefix\doi{10.1137/080727890}.
\bibitem[{Stumm and Walther(2010)}]{stumm2010}
\bibinfo{author}{P.~Stumm}, \bibinfo{author}{A.~Walther},
\newblock \bibinfo{title}{New algorithms for optimal online checkpointing},
\newblock \bibinfo{journal}{SIAM Journal on Scientific Computing}
  \bibinfo{volume}{32} (\bibinfo{year}{2010}) \bibinfo{pages}{836--854}.
  \DOIprefix\doi{10.1137/080742439}.
\bibitem[{Stumm and Walther(2009)}]{stumm2009}
\bibinfo{author}{P.~Stumm}, \bibinfo{author}{A.~Walther},
\newblock \bibinfo{title}{{MultiStage} approaches for optimal offline
  checkpointing},
\newblock \bibinfo{journal}{SIAM Journal on Scientific Computing}
  \bibinfo{volume}{31} (\bibinfo{year}{2009}) \bibinfo{pages}{1946--1967}.
  \DOIprefix\doi{10.1137/080718036}.
\bibitem[{Aupy et~al.(2016)Aupy, Herrmann, Hovland, and Robert}]{aupy2016}
\bibinfo{author}{G.~Aupy}, \bibinfo{author}{J.~Herrmann},
  \bibinfo{author}{P.~Hovland}, \bibinfo{author}{Y.~Robert},
\newblock \bibinfo{title}{Optimal multistage algorithm for adjoint
  computation},
\newblock \bibinfo{journal}{SIAM Journal on Scientific Computing}
  \bibinfo{volume}{38} (\bibinfo{year}{2016}) \bibinfo{pages}{C232--C255}.
  \DOIprefix\doi{10.1137/15M1019222}.
\bibitem[{Schanen et~al.(2016)Schanen, Marin, Zhang, and
  Anitescu}]{schanen2016}
\bibinfo{author}{M.~Schanen}, \bibinfo{author}{O.~Marin},
  \bibinfo{author}{H.~Zhang}, \bibinfo{author}{M.~Anitescu},
\newblock \bibinfo{title}{Asynchronous two-level checkpointing scheme for
  large-scale adjoints in the spectral-element solver {N}ek5000},
\newblock \bibinfo{journal}{Procedia Computer Science} \bibinfo{volume}{80}
  (\bibinfo{year}{2016}) \bibinfo{pages}{1147--1158}.
  \DOIprefix\doi{10.1016/j.procs.2016.05.444}.
\bibitem[{Herrmann and {Pallez (Aupy)}(2020)}]{herrmann2020}
\bibinfo{author}{J.~Herrmann}, \bibinfo{author}{G.~{Pallez (Aupy)}},
\newblock \bibinfo{title}{{H-Revolve}: a framework for adjoint computation on
  synchronous hierarchical platforms},
\newblock \bibinfo{journal}{ACM Transactions on Mathematical Software}
  \bibinfo{volume}{46} (\bibinfo{year}{2020}). \DOIprefix\doi{10.1145/3378672}.
\bibitem[{Kukreja et~al.(2019)Kukreja, H\"{u}ckelheim, Louboutin, Hovland, and
  Gorman}]{kukreja2019}
\bibinfo{author}{N.~Kukreja}, \bibinfo{author}{J.~H\"{u}ckelheim},
  \bibinfo{author}{M.~Louboutin}, \bibinfo{author}{P.~Hovland},
  \bibinfo{author}{G.~Gorman},
\newblock \bibinfo{title}{Combining checkpointing and data compression to
  accelerate adjoint-based optimization problems},
\newblock in: \bibinfo{editor}{R.~Yahyapour} (Ed.),
  \bibinfo{booktitle}{Euro-Par 2019: Parallel Processing},
  \bibinfo{publisher}{Springer Nature Switzerland AG}, \bibinfo{year}{2019},
  pp. \bibinfo{pages}{87--100}.
\bibitem[{Naumann(2008)}]{naumann2008}
\bibinfo{author}{U.~Naumann},
\newblock \bibinfo{title}{Optimal {J}acobian accumulation is {NP}-complete},
\newblock \bibinfo{journal}{Mathematical Programming} \bibinfo{volume}{112}
  (\bibinfo{year}{2008}) \bibinfo{pages}{427--441}.
  \DOIprefix\doi{10.1007/s10107-006-0042-z}.
\bibitem[{Jain et~al.(2020)Jain, Jain, Nrusimha, Gholami, Abbeel, Gonzalez,
  Keutzer, and Stoica}]{jain2020}
\bibinfo{author}{P.~Jain}, \bibinfo{author}{A.~Jain},
  \bibinfo{author}{A.~Nrusimha}, \bibinfo{author}{A.~Gholami},
  \bibinfo{author}{P.~Abbeel}, \bibinfo{author}{J.~Gonzalez},
  \bibinfo{author}{K.~Keutzer}, \bibinfo{author}{I.~Stoica},
\newblock \bibinfo{title}{Checkmate: Breaking the memory wall with optimal
  tensor rematerialization},
\newblock in: \bibinfo{editor}{I.~Dhillon},
  \bibinfo{editor}{D.~Papailiopoulos}, \bibinfo{editor}{V.~Sze} (Eds.),
  \bibinfo{booktitle}{Proceedings of Machine Learning and Systems},
  volume~\bibinfo{volume}{2}, \bibinfo{year}{2020}, pp.
  \bibinfo{pages}{497--511}.
\bibitem[{Kirisame et~al.(2021)Kirisame, Lyubomirsky, Haan, Brennan, He,
  Roesch, Chen, and Tatlock}]{kirisame2021-preprint}
\bibinfo{author}{M.~Kirisame}, \bibinfo{author}{S.~Lyubomirsky},
  \bibinfo{author}{A.~Haan}, \bibinfo{author}{J.~Brennan},
  \bibinfo{author}{M.~He}, \bibinfo{author}{J.~Roesch},
  \bibinfo{author}{T.~Chen}, \bibinfo{author}{Z.~Tatlock},
\newblock \bibinfo{title}{Dynamic tensor rematerialization}
  (\bibinfo{year}{2021}). \URLprefix \url{https://arxiv.org/abs/2006.09616v4}.
  \href{http://arxiv.org/abs/2006.09616v4}{{\tt arXiv:2006.09616v4}}.
\bibitem[{Hasco\"{e}t et~al.(2005)Hasco\"{e}t, Naumann, and
  Pascual}]{hascoet2005}
\bibinfo{author}{L.~Hasco\"{e}t}, \bibinfo{author}{U.~Naumann},
  \bibinfo{author}{V.~Pascual},
\newblock \bibinfo{title}{``{T}o be recorded'' analysis in reverse-mode
  automatic differentiation},
\newblock \bibinfo{journal}{Future Generation Computer Systems}
  \bibinfo{volume}{21} (\bibinfo{year}{2005}) \bibinfo{pages}{1401--1417}.
  \DOIprefix\doi{10.1016/j.future.2004.11.009}.
\bibitem[{Zhang and Constantinescu(2021)}]{zhang2021}
\bibinfo{author}{H.~Zhang}, \bibinfo{author}{E.~Constantinescu},
\newblock \bibinfo{title}{Revolve-based adjoint checkpointing for multistage
  time integration},
\newblock in: \bibinfo{editor}{M.~Paszynski},
  \bibinfo{editor}{D.~Kranzlm\"{u}ller}, \bibinfo{editor}{V.~V.
  Krzhizhanovskaya}, \bibinfo{editor}{J.~J. Dongarra},
  \bibinfo{editor}{P.~M.~A. Sloot} (Eds.), \bibinfo{booktitle}{Computational
  Science -- ICCS 2021}, \bibinfo{publisher}{Springer Nature Switzerland AG},
  \bibinfo{year}{2021}, pp. \bibinfo{pages}{451--464}.
\bibitem[{Zhang and Constantinescu(2023)}]{zhang2023}
\bibinfo{author}{H.~Zhang}, \bibinfo{author}{E.~M. Constantinescu},
\newblock \bibinfo{title}{Optimal checkpointing for adjoint multistage
  time-stepping schemes},
\newblock \bibinfo{journal}{Journal of Computational Science}
  \bibinfo{volume}{66} (\bibinfo{year}{2023}) \bibinfo{pages}{101913}.
  \DOIprefix\doi{10.1016/j.jocs.2022.101913}.
\bibitem[{Maddison et~al.(2019)Maddison, Goldberg, and Goddard}]{maddison2019}
\bibinfo{author}{J.~R. Maddison}, \bibinfo{author}{D.~N. Goldberg},
  \bibinfo{author}{B.~D. Goddard},
\newblock \bibinfo{title}{Automated calculation of higher order partial
  differential equation constrained derivative information},
\newblock \bibinfo{journal}{SIAM Journal on Scientific Computing}
  \bibinfo{volume}{41} (\bibinfo{year}{2019}) \bibinfo{pages}{C417--C445}.
  \DOIprefix\doi{10.1137/18M1209465}.
\bibitem[{Pringle et~al.(2016)Pringle, Jones, Goswami, Narayanan, and
  Goldberg}]{pringle2016}
\bibinfo{author}{G.~J. Pringle}, \bibinfo{author}{D.~C. Jones},
  \bibinfo{author}{S.~Goswami}, \bibinfo{author}{S.~H.~K. Narayanan},
  \bibinfo{author}{D.~Goldberg}, \bibinfo{title}{Providing the {ARCHER}
  community with adjoint modelling tools for high-performance oceanographic and
  cryospheric computation}, \bibinfo{type}{Technical Report}, EPCC,
  \bibinfo{year}{2016}. \bibinfo{note}{Version 1.1}.
\bibitem[{Goldberg et~al.(2020)Goldberg, Smith, Narayanan, Heimbach, and
  Morlighem}]{goldberg2020}
\bibinfo{author}{D.~N. Goldberg}, \bibinfo{author}{T.~A. Smith},
  \bibinfo{author}{S.~H.~K. Narayanan}, \bibinfo{author}{P.~Heimbach},
  \bibinfo{author}{M.~Morlighem},
\newblock \bibinfo{title}{Bathymetric influences on {A}ntarctic ice-shelf melt
  rates},
\newblock \bibinfo{journal}{Journal of Geophysical Research: Oceans}
  \bibinfo{volume}{125} (\bibinfo{year}{2020}) \bibinfo{pages}{e2020JC016370}.
  \DOIprefix\doi{10.1029/2020JC016370}.
\bibitem[{Rathgeber et~al.(2016)Rathgeber, Ham, Mitchell, Lange, Luporini,
  Mcrae, Bercea, Markall, and Kelly}]{rathgeber2016}
\bibinfo{author}{F.~Rathgeber}, \bibinfo{author}{D.~A. Ham},
  \bibinfo{author}{L.~Mitchell}, \bibinfo{author}{M.~Lange},
  \bibinfo{author}{F.~Luporini}, \bibinfo{author}{A.~T.~T. Mcrae},
  \bibinfo{author}{G.-T. Bercea}, \bibinfo{author}{G.~R. Markall},
  \bibinfo{author}{P.~H.~J. Kelly},
\newblock \bibinfo{title}{Firedrake: Automating the finite element method by
  composing abstractions},
\newblock \bibinfo{journal}{ACM Transactions on Mathematical Software}
  \bibinfo{volume}{43} (\bibinfo{year}{2016}). \DOIprefix\doi{10.1145/2998441}.
\bibitem[{Dolci et~al.(2024)Dolci, Maddison, Ham, Pallez, and
  Herrmann}]{dolci2024}
\bibinfo{author}{D.~I. Dolci}, \bibinfo{author}{J.~R. Maddison},
  \bibinfo{author}{D.~A. Ham}, \bibinfo{author}{G.~Pallez},
  \bibinfo{author}{J.~Herrmann},
\newblock \bibinfo{title}{{c}heckpoint\_schedules: schedules for incremental
  checkpointing of adjoint simulations},
\newblock \bibinfo{journal}{The Journal of Open Source Software}
  \bibinfo{volume}{9} (\bibinfo{year}{2024}) \bibinfo{pages}{6148}.
  \DOIprefix\doi{10.21105/joss.06148}.
\bibitem[{Mitusch(2018)}]{mitusch2018}
\bibinfo{author}{S.~K. Mitusch}, \bibinfo{title}{An algorithmic differentiation
  tool for {FEniCS}}, Master's thesis, University of Oslo,
  \bibinfo{year}{2018}.
\bibitem[{Abadi et~al.(2016)Abadi, Barham, Chen, Chen, Davis, Dean, Devin,
  Ghemawat, Irving, Isard, Kudlur, Levenberg, Monga, Moore, Murray, Steiner,
  Tucker, Vasudevan, Warden, Wicke, Yu, and Zhen}]{abadi2016}
\bibinfo{author}{M.~Abadi}, \bibinfo{author}{P.~Barham},
  \bibinfo{author}{J.~Chen}, \bibinfo{author}{Z.~Chen},
  \bibinfo{author}{A.~Davis}, \bibinfo{author}{J.~Dean},
  \bibinfo{author}{M.~Devin}, \bibinfo{author}{S.~Ghemawat},
  \bibinfo{author}{G.~Irving}, \bibinfo{author}{M.~Isard},
  \bibinfo{author}{M.~Kudlur}, \bibinfo{author}{J.~Levenberg},
  \bibinfo{author}{R.~Monga}, \bibinfo{author}{S.~Moore},
  \bibinfo{author}{D.~G. Murray}, \bibinfo{author}{B.~Steiner},
  \bibinfo{author}{P.~Tucker}, \bibinfo{author}{V.~Vasudevan},
  \bibinfo{author}{P.~Warden}, \bibinfo{author}{M.~Wicke},
  \bibinfo{author}{Y.~Yu}, \bibinfo{author}{X.~Zhen},
\newblock \bibinfo{title}{{TensorFlow}: a system for large-scale machine
  learning},
\newblock in: \bibinfo{booktitle}{12th USENIX Symposium on Operating Systems
  Design and Implementation (OSDI '16)}, \bibinfo{year}{2016}, pp.
  \bibinfo{pages}{265--283}.
\bibitem[{Vallis(2006)}]{vallis2006}
\bibinfo{author}{G.~K. Vallis}, \bibinfo{title}{Atmospheric and oceanic fluid
  dynamics}, \bibinfo{publisher}{Cambridge University Press},
  \bibinfo{year}{2006}. \bibinfo{note}{Third printing 2008}.
\bibitem[{Lam et~al.(2015)Lam, Pitrou, and Seibert}]{lam2015}
\bibinfo{author}{S.~K. Lam}, \bibinfo{author}{A.~Pitrou},
  \bibinfo{author}{S.~Seibert},
\newblock \bibinfo{title}{Numba: a {LLVM}-based {P}ython {JIT} compiler},
\newblock in: \bibinfo{booktitle}{LLVM '15: Proceedings of the Second Workshop
  on the LLVM Compiler Infrastructure in HPC}, \bibinfo{publisher}{Association
  for Computing Machinery}, \bibinfo{year}{2015}, pp. \bibinfo{pages}{1--6}.
  \DOIprefix\doi{10.1145/2833157.2833162}.
\bibitem[{fir(2024)}]{firedrake_a}
\bibinfo{title}{Software used in `{O}n the implementation of checkpointing with
  high-level algorithmic differentiation'}, \bibinfo{howpublished}{Zenodo},
  \bibinfo{year}{2024}. \DOIprefix\doi{10.5281/zenodo.11104674}.
\bibitem[{tlm(2024)}]{tlm_adjoint_c}
\bibinfo{title}{tlm-adjoint/tlm\_adjoint: tlm\_adjoint 2024-05-06},
  \bibinfo{howpublished}{Zenodo}, \bibinfo{year}{2024}.
  \DOIprefix\doi{10.5281/zenodo.7695474}.
\bibitem[{Maddison(2024)}]{stommel_benchmark_b}
\bibinfo{author}{J.~R. Maddison}, \bibinfo{title}{Benchmark scripts for
  `{S}tep-based checkpointing with high-level algorithmic differentiation'},
  \bibinfo{howpublished}{Zenodo}, \bibinfo{year}{2024}.
  \DOIprefix\doi{10.5281/zenodo.11149901}.
\bibitem[{Stommel(1948)}]{stommel1948}
\bibinfo{author}{H.~Stommel},
\newblock \bibinfo{title}{The westward intensification of wind-driven ocean
  currents},
\newblock \bibinfo{journal}{Eos, Transactions American Geophysical Union}
  \bibinfo{volume}{29} (\bibinfo{year}{1948}) \bibinfo{pages}{202--206}.
  \DOIprefix\doi{10.1029/TR029i002p00202}.
\bibitem[{Balay et~al.(2023)Balay, Abhyankar, Adams, Benson, Brown, Brune,
  Buschelman, Constantinescu, Dalcin, Dener, Eijkhout, Faibussowitsch, Gropp,
  Hapla, Isaac, Jolivet, Karpeev, Kaushik, Knepley, Kong, Kruger, May, McInnes,
  Mills, Mitchell, Munson, Roman, Rupp, Sanan, Sarich, Smith, Zampini, Zhang,
  Zhang, and Zhang}]{petsc-user-3.20}
\bibinfo{author}{S.~Balay}, \bibinfo{author}{S.~Abhyankar},
  \bibinfo{author}{M.~F. Adams}, \bibinfo{author}{S.~Benson},
  \bibinfo{author}{J.~Brown}, \bibinfo{author}{P.~Brune},
  \bibinfo{author}{K.~Buschelman}, \bibinfo{author}{E.~Constantinescu},
  \bibinfo{author}{L.~Dalcin}, \bibinfo{author}{A.~Dener},
  \bibinfo{author}{V.~Eijkhout}, \bibinfo{author}{J.~Faibussowitsch},
  \bibinfo{author}{W.~D. Gropp}, \bibinfo{author}{V.~Hapla},
  \bibinfo{author}{T.~Isaac}, \bibinfo{author}{P.~Jolivet},
  \bibinfo{author}{D.~Karpeev}, \bibinfo{author}{D.~Kaushik},
  \bibinfo{author}{M.~G. Knepley}, \bibinfo{author}{F.~Kong},
  \bibinfo{author}{S.~Kruger}, \bibinfo{author}{D.~A. May},
  \bibinfo{author}{L.~C. McInnes}, \bibinfo{author}{R.~T. Mills},
  \bibinfo{author}{L.~Mitchell}, \bibinfo{author}{T.~Munson},
  \bibinfo{author}{J.~E. Roman}, \bibinfo{author}{K.~Rupp},
  \bibinfo{author}{P.~Sanan}, \bibinfo{author}{J.~Sarich},
  \bibinfo{author}{B.~F. Smith}, \bibinfo{author}{S.~Zampini},
  \bibinfo{author}{H.~Zhang}, \bibinfo{author}{H.~Zhang},
  \bibinfo{author}{J.~Zhang}, \bibinfo{title}{{PETSc/TAO} Users Manual},
  \bibinfo{type}{Technical Report} \bibinfo{number}{ANL-21/39 - Revision 3.20},
  \bibinfo{year}{2023}. \DOIprefix\doi{10.2172/2205494}.

\end{thebibliography}

\end{document}